\documentclass[12pt]{article}
\usepackage{graphicx}
\usepackage{epstopdf}
\usepackage{amsthm}
\usepackage{amsmath}
\usepackage{amssymb}
\usepackage{amstext}
\usepackage{amsfonts}
\usepackage{enumerate}
\usepackage[authoryear,nonamebreak,bibstyle]{natbib}
\usepackage{footnote}
\makesavenoteenv{tabular}
\makesavenoteenv{table}
\usepackage[shortlabels]{enumitem}
\usepackage{multirow}
\usepackage[title, toc, page]{appendix}
\usepackage{xcolor}
\usepackage{setspace}
\usepackage{cancel}
\usepackage{soul}

\usepackage{sectsty}
\subsubsectionfont{\normalfont\itshape}

\begin{document}
\baselineskip=.22in\parindent=30pt

\newtheorem{tm}{Theorem}
\newtheorem{dfn}{Definition}
\newtheorem{lma}{Lemma}
\newtheorem{assu}{Assumptions}
\newtheorem*{assum}{Assumptions}
\newtheorem{prop}{Proposition}
\newtheorem{cro}{Corollary}
\newtheorem{example}{Example}
\newcommand{\exm}{\begin{example}}
\newcommand{\exmm}{\end{example}}
\newtheorem*{theorem*}{Theorem}
\newcommand{\cor}{\begin{cro}}
\newcommand{\corr}{\end{cro}}
\newtheorem{exa}{Example}
\newcommand{\ex}{\begin{exa}}
\newcommand{\exx}{\end{exa}}
\newtheorem{remak}{Remark}
\newcommand{\rmk}{\begin{remak}}
\newcommand{\rmkk}{\end{remak}}
\newcommand{\thm}{\begin{tm}}
\newcommand{\nt}{\noindent}
\newcommand{\thmm}{\end{tm}}
\newcommand{\lm}{\begin{lma}}
\newcommand{\lmm}{\end{lma}}
\newcommand{\ass}{\begin{assu}}
\newcommand{\asss}{\end{assu}}
\newcommand{\assm}{\begin{assum}}
\newcommand{\assmm}{\end{assum}}
\newcommand{\df}{\begin{dfn}  }
\newcommand{\dff}{\end{dfn}}
\newcommand{\prp}{\begin{prop}}
\newcommand{\prpp}{\end{prop}}
\newcommand{\bqu}{\sloppy \small \begin{quote}}
\newcommand{\equ}{\end{quote} \sloppy \large}
\newcommand\cites[1]{\citeauthor{#1}'s\ (\citeyear{#1})}

\newcommand{\eq}{\begin{equation}}
\newcommand{\eqq}{\end{equation}}
\newtheorem{claim}{\it Claim}
\newcommand{\cl}{\begin{claim}}
\newcommand{\cll}{\end{claim}}
\newcommand{\bit}{\begin{itemize}}
\newcommand{\eit}{\end{itemize}}
\newcommand{\ben}{\begin{enumerate}}
\newcommand{\een}{\end{enumerate}}
\newcommand{\bcen}{\begin{center}}
\newcommand{\ecen}{\end{center}}
\newcommand{\fn}{\footnote}
\newcommand{\ds}{\begin{description}}
\newcommand{\dss}{\end{description}}
\newcommand{\prf}{\begin{proof}}
\newcommand{\prff}{\end{proof}}
\newcommand{\cs}{\begin{cases}}
\newcommand{\css}{\end{cases}}
\newcommand{\ml}{\item}
\newcommand{\lb}{\label}
\newcommand{\ra}{\rightarrow}
\newcommand{\tra}{\twoheadrightarrow}
\newcommand*{\supp}{\operatornamewithlimits{sup}\limits}
\newcommand*{\inff}{\operatornamewithlimits{inf}\limits}
\newcommand{\nf}{\normalfont}
\renewcommand{\Re}{\mathbb{R}}
\newcommand{\Ze}{\mathbb Z}
\newcommand*{\mmax}{\operatornamewithlimits{max}\limits}
\newcommand*{\mmin}{\operatornamewithlimits{min}\limits}
\newcommand*{\argmax}{\operatornamewithlimits{arg max}\limits}
\newcommand*{\argmin}{\operatornamewithlimits{arg min}\limits}

\newcommand{\CR}{\mathcal R}
\newcommand{\CC}{\mathcal C}
\newcommand{\CT}{\mathcal T}

\newtheorem{innercustomthm}{Theorem}
\newenvironment{customthm}[1]
  {\renewcommand\theinnercustomthm{#1}\innercustomthm}
  {\endinnercustomthm}
\newtheorem{einnercustomthm}{Extended Theorem}
\newenvironment{ecustomthm}[1]
  {\renewcommand\theeinnercustomthm{#1}\einnercustomthm}
  {\endeinnercustomthm}

\newcommand{\red}{\textcolor{red}}
\newcommand{\blue}{\textcolor{blue}}
\newcommand{\purple}{\textcolor{purple}}
\newcommand{\mred}[1]{\color{red}{#1}\color{black}}
\newcommand{\mblue}[1]{\color{blue}{#1}\color{black}}
\newcommand{\mpurple}[1]{\color{purple}{#1}\color{black}}
\makeatletter
\newcommand{\customlabel}[2]{%
\protected@write \@auxout {}{\string \newlabel {#1}{{#2}{}}}}
\makeatother

\def\qed{\hfill\vrule height4pt width4pt
depth0pt}
\def\reff #1\par{\noindent\hangindent =\parindent
\hangafter =1 #1\par}
\def\title #1{\begin{center}
{\Large {\bf #1}}
\end{center}}
\def\author #1{\begin{center} {\large #1}
\end{center}}
\def\date #1{\centerline {\large #1}}
\def\place #1{\begin{center}{\large #1}
\end{center}}

\def\date #1{\centerline {\large #1}}
\def\place #1{\begin{center}{\large #1}\end{center}}
\def\intr #1{\stackrel {\circ}{#1}}
\def\R{{\rm I\kern-1.7pt R}}
 \def\N{{\rm I}\hskip-.13em{\rm N}}
 \newcommand{\cprod}{\Pi_{i=1}^\ell}
\let\Large=\large
\let\large=\normalsize

\begin{titlepage}
\def\thefootnote{\fnsymbol{footnote}}
\vspace*{1.1in}

\title{Topological Connectedness and Behavioral Assumptions  on \\ \vskip .5em Preferences: A Two-Way Relationship\fn{A previous version of this paper was completed while Uyan{\i}k was a post-doctoral fellow at the {\it Allen Wallis Institute of Political Economy} at the University of Rochester;   earlier versions  were  presented at     Rutgers University (April 19, 2017), the University of Rochester (May 4, 2017), the University of Queensland  (August 8, 2018),  the {\it 17th SAET Conference} in Faro (June 30, 2017), and  
 the {\it 36th Australasian Economic Theory Workshop} in Canberra  (February 8, 2018).  We are grateful to Rich McLean for the SAET invitation, and thank him for detailed and indispensable advice on the presentation of our results. We also thank Paul Anand,  Han Bleichrodt, Ying Chen, John Duggan,  H\"ulya Eraslan,  Filippo Massari,  Andy McLennan, Onur Ozgur,  John Quah,  Eddie Schlee and Rajiv Vohra for encouragement and stimulating conversation. }} 

 
\author{M. Ali Khan\fn{Department of Economics, Johns Hopkins University, Baltimore, MD 21218.} and  Metin Uyan{\i}k\footnote{School of Economics, University of Queensland, Brisbane, QLD 4072.}}

\vskip 1.00em
\date{October 25, 2018}

\vskip 1.75em

\vskip 1.00em

\baselineskip=.18in

\noindent{\bf Abstract:} 
This paper offers a comprehensive treatment of the question 
as to whether a binary relation can be consistent (transitive) without being decisive (complete), or decisive without being consistent, or simultaneously inconsistent or indecisive, in the presence of a    continuity hypothesis that is, in principle, non-testable.   It  identifies topological connectedness of the (choice) set over which the continuous binary relation is defined as being crucial to this question. Referring to  the two-way relationship as the Eilenberg-Sonnenschein (ES)  research program, it presents four synthetic, and complete, characterizations of connectedness,   and its natural extensions;  and two consequences that only stem from it.    The six theorems  are  novel to both the  economic and the mathematical literature: they     generalize  pioneering results of \cite{ei41}, \cite{so65}, \cite{sc71} and \cite{se69}, and  are  relevant  to  several applied contexts, as well as  to ongoing theoretical work.    \hfill  (140~words)

\vskip 1.25em

\vskip 1.75em

\noindent {\it Journal of Economic Literature} Classification
Numbers: C00, D00, D01

\vskip .65em

\noindent {\it 2010 Mathematics Subject} Classification Numbers: 91B55, 37E05.

\vskip .65em


\vskip .65em

\noindent {\it Running Title:}  Topological Connectedness and Preferences

\end{titlepage}

\tableofcontents




\pagebreak 

\setcounter{footnote}{0}


\setlength{\abovedisplayskip}{0.02cm}
\setlength{\belowdisplayskip}{0.02cm}


\bqu Just as it is possible to speak prose without noticing that fact ...., it is possible also to be talking topology without sensing any topology around us. 

The fear of being \lq unmeasurable' can be a rather raw worry (more polemical than illuminating) and such a diagnosis can serve as a reactionary diversion from reasoning (reasoning that we can sensibly use). ...  [it]  explains why set theory (and, based on that, topological ideas) can be very useful in practical  economic and social evaluation.\fn{The first sentence is taken from p. 298, and the second from p. 367 in \cite{se17}. Sen fills the ellipsis in the first by referring to Moli\`ere, and in the second, by the Aristotelian counsel that one aspire to as much precision as the subject deserves. This is also the place to acknowledge our substantial indebtedness to \cite{so65} and \cite{se17} for rekindling our interest in the subject on which we report in this paper: in fact, the conclusion of the former could be read as spelling out  the basic motivation for this paper as well.    } \hfill{Sen~(2017)}  \equ

\section{Introduction} \lb{sec: introduction}  
The postulates of transitivity, completeness and continuity of a binary relation defined over a (choice) set are basic to modern microeconomic theory, and this paper addresses itself to the logical implications of these properties in an exclusively  topological register.    
In particular, it asks whether an agent can be {\it decisive,} in the formal sense of having complete preferences, without being {\it consistent,} again  in the  formal sense of having transitive preferences?  Or to turn the matter on its head, whether she can be consistent without being decisive? Or even, in the case of an anti-symmetric binary relation, inconsistent or indecisive or both?  It shows, to put it in a nutshell, that all these possibilities  are  foreclosed  under the standard technical assumption of continuity (typically taken to be an innocuous technical property that cannot be falsified by any finite number of observations) in any  topology in which the choice set is connected. Indeed, the paper  goes  beyond these three basic questions to consider   additional complementary queries, and  in their turn,   settles them  by appealing to  the interplay of the topological properties of continuity and connectedness.  As such, the  results   raise a subtle but striking challenge to the modeling of agency in the theory of binary choice. 

To be sure, the questions we ask have a venerable history in economic theory and in mathematics, though they have not been posed and investigated in the way that we do  here. The treatment in the antecedent literature has been piecemeal.\fn{See, for example, \citet{bb82}, a survey that still remains fresh and worth reading. The authors cite Eilenberg,  Sonnenschein and Schmeidler, reproduce the proof of Schmeidler's result, differentiate points of departure based on whether strict or weak preferences as used as primitive, but do not make the connections that we do here.   The properties of the binary relation referred to below constitute  standard textbook material, as in \cite{bb82} or \cite{se17} for example, but  precise definitions are spelt out for the reader's convenience in the   section to follow.   \label{fn:bb}}   Thus, in the context of a connected choice set, and a correspondingly continuous binary relation,
\ben[topsep=5pt]
\setlength{\itemsep}{1pt} 
\item \cite{ei41} showed that completeness and the impossibility of  indifference between  any two distinct items (a sort of {\it extreme decisiveness}),  imply transitivity,  \item \cite{so65} showed that completeness and  semi-transitivity of the binary relation  imply its transitivity,\fn{We may point out here that \cite{so65} is an elaboration of his Ph.D.\,dissertation that dispensed with Eilenberg's requirement  that the binary relation be anti-symmetric, that  preferences not be  
 constrained by the requirement of   singleton indifference sets. We return to this point in the sequel. In fact, a careful reader would have already noted Sonnenschein making this point in the very first paragraph of his paper.}  and  
 \item  \cite{sc71} showed that  transitivity and non-triviality of the binary relation imply its completeness.   \een
\cite{ei41} also furnished necessary and sufficient conditions for a connected topological space to be {\it ordered,} which is to say, conditions under which  it admits an  anti-symmetric, transitive,  complete and continuous relation on it.\fn{We return to this theorem below. It is perhaps worth reminding the  reader that  Nachbin's pioneering monograph is titled  \lq\lq Topology and Order;" see \cite{wa54} and \cite{bm95} for details.  \label{fn:ei}}

Connected topologies are of course ubiquitous in economic theory,\fn{We need hardly remind the reader that finite-dimensional Euclidean spaces, the simplex of (objective) lotteries over a finite set, and infinite-dimensional topological vector spaces are some of the most obvious examples.}   and these pioneering contributions testify to the substantive restrictions they engender when the  continuity property of the ambient  binary relation is indexed by them.   The  sufficiency theorems they offer   move {\it forward} from connectedness, and they all question the opinion, still pervasive in some circles,  that topological assumptions on the choice set, and on the objectives that are  defined over that set, have no substantive and behavioral content. They are  simply seen as innocuous technical requirements made to guarantee the existence of an optimal choice.\fn{Some would say that this  is almost professional folk-wisdom still: see for example page 80 (last paragraph) in \cite{gi09}.}  Thus \cite{wa88a} writes: 
\bqu
[...] by themselves these topological assumptions are merely \lq\lq technical." They have no empirical implications and cannot be verified or falsified by observations. Technical assumptions, while not very satisfactory, are not very bothersome either. They merely serve to make mathematical machinery work smoothly. They are void of empirical meaning, and so do not entail obscurity.\fn{By truncating Wakker's words as we do, we take them out of context; we return to them, and to  \cite{na85},   in Section 4 below.  \label{fn:wa}} 
\equ

 \nt It was perhaps Schmeidler who offered, though more understatedly than warranted,  the sharpest counter to this prevailing opinion. He   
  wrote:

\bqu Order properties of preference relations have intuitive meaning in the context of the behavioral sciences. This is not the case with topological conditions. They are only assumed in order to utilize the mathematical tools applied in the analysis of some problems in the behavioral sciences. They may imply, however, as in the theorem, a very restrictive condition of plausible nature. \equ 

\noindent In an article scarcely over a page long, and without any references to its  precursors in the work not only  of Eilenberg and  Sonnenschein, but also \cite{ra63}  and \cite{se69}, Schmeidler  presented a theorem that  confined  itself solely to the topological register.   With all linear structures, in particular,  dispensed with, he articulated the claim  that  any topology that ensured   connectedness of the choice set, and the continuity of a non-trivial and transitive relation over that set, necessarily guarantees the completeness of the relation -- a sharp and influential example of a technical requirement leading to a substantive  conclusion of a behavioral nature. 

However, one can ask 
 whether Schmeidler's  results, as well as those of the others,     can be framed under a broader rubric that enables one to see how  one   result can be reinforced by another, a mutual  strengthening that offers a synthetic and more comprehensive overview. Put differently,  can one ask  for necessity results that move {\it backward} to connectedness? Indeed, inspite of his focus on the question of a numerical representation of a given preference relation over the choice space,\fn{This aspect of Eilenberg's paper was taken up by \cite{de54} and its extensive study now constitutes a rather rich history to which we shall briefly refer to in the sequel. \label{fn:rep}}   Eilenberg had already furnished necessary and sufficient conditions for  the existence of  topology that provided a simultaneous  guarantee both of the connectedness of the choice set and the continuity of the relation defined over the set. And  Sonnenschein's work can be seen as a further  elaboration of this line of inquiry to a setting  in which     binary relations are  unconstrained by Eilenberg's  singleton indifference sets, a context of primary interest for microeconomic theory.  And surely, the next step in
the line of development is  Sen's. Even though his primary emphasis is on choice functions and problems of social choice, his forensic examination of the transitivity postulate, one that abstains from topological considerations,  draws on Eilenberg, Sonnenschein and  Lorimer as a subtext, and  thereby leaves open the question as to the consequences of putting  continuity considerations back in.\fn{There is some irony in that we extend Sen's work in a direction that he explicitly wanted to avoid. He criticizes the literature's \lq\lq  overwhelming concern with transitivity and continuity," and  referring to his results, writes how they are  \lq\lq derived without any assumption
 of continuity and related conditions as these conditions might not be very realistic for
 certain problems of rational choice, especially in dealing with social decisions."  \label{fn:sen}}
 In any case, a principal motivation behind this paper is to focus on this simultaneity:  an  inquiry that moves forward and backward in its investigation of not only  what  topological connectedness  implies for a  binary relation continuous in that topology, but also what   is  implied in its turn by the substantive properties of that class of relations. It is to focus on both directions of a two-way relationship between  topological and order structures. As such, we refer to this investigation  as the Eilenberg-Sonnenschein (ES) research program, recognizing that even though it is the forward (sufficiency) direction that is of primary interest to economists, and the backward (necessary) direction of primary interest to topologists, the four equivalence theorems that we present below are perhaps two sides of the same coin -- a productive two-way relationship in which assumptions on the choice set, and  the objectives that are defined on that set,  have strong and obvious implications for each other. Eilenberg  saw this more than seventy-five years ago.

In summary, without being overly pompous and pedantic, and motivated primarily to make an exploratory essay in pure theory\fn{We use the phrase \lq\lq pure theory" in the Marshallian sense of the \lq\lq The pure theory of international trade."} reader-friendly, we list what we see  to be the six   contributions of this paper, and leave it to the reader to   separate the primary from the secondary.  
\ben[topsep=5pt]
\setlength{\itemsep}{1pt} \item Gathering and connecting under one rubric the  three foundational\fn{This is nothing but a loose summary sentence of the ideas explicated in the four paragraphs above. We shall reserve the term \lq foundational' for the results enumerated in the second paragraph above. }  results enumerated above on how continuity and connectedness in a given topology logically link  transitivity and completeness. \item Remaining within the parametric ambit of these foundational results,   a generalization of Eilenberg's theorem that non-triviality and no-indifference between distinct items imply both transitivity and completeness.\fn{For a precise statement, see 
(a)$\Rightarrow$(b) in Theorem 1, and its detailed discussion encapsulated in Proposition 1 below.  This may also be the place to note that  Eilenberg's equivalence result is not parallel to ours and of a different  nature.}  \item   Turning the foundational results around by asking for the type of topologies under which the logical implications on the binary relations hold, and providing complete answers. \item  Generalizing the collectivity of these results by deconstructing notions of transitivity, completeness and   connectedness, and establishing the tightness of these generalizations. \item Investigating ancillary notions  such as {\it fragility, flimsiness} and {\it separability} that fall within the ES program, even though not specifically related to transitivity and completeness. \item Drawing the  implications of the results for  a variety of applied contexts under the headings of {\it redundancy} and {\it hiddenness}.   \een

Once the basic motivation of the paper is understood, and the results outlined,  the presentation of the rest of the paper follows rather naturally. Section 2 develops the basic notation and vocabulary of the subject  and  Section 3 presents the results in four subsections:  the four equivalence theorems in the first three, and a sufficiency theorem that casts  the results of Sen and Sonnenschein in the framework of incomplete preferences. The latter, and its affiliated proposition, while of interest for its own sake,  also  undergirds the proofs of the  equivalence theorems that precede it. 
Section 4 is devoted to a somewhat hurried discussion and application of the results to a variety of applied registers.  Section 5  recollects the strands already laid out in this introduction, and concludes with two  observations for further work. Section 6
is devoted to the proofs.

\section{Notational and Conceptual Preliminaries } \lb{sec: mp}
Let $X$ be a set and a {\it binary relation} $R$ on it as a subset $R\subset X\times X.$ We define  
\[
\begin{array}{lcl}
R(x)&=&\{y~|~ (x,y)\in R\},\\
R^{-1}(x)&=&\{y~|~ (y,x)\in R\},\\
R^{-1}&=&\{(x,y)~|~(y,x)\in R\},
\end{array}
\]
where  $R(x)$ denotes the {\it upper section} of $R$ at $x,$ $R^{-1}(x)$ the {\it lower section} of $R$ at $x$ and $R^{-1}$ the {\it transpose} of $R.$ Let $\Delta=\{(x,x)|x\in X\}$ and $R^c$  the complement of $R.$  We say that $R$ has {\it open {\nf (}closed{\nf )} sections} if its upper and lower sections are open (closed) in the topology that $X$ is endowed with.  We call $R$ {\it continuous} if its sections are closed and the sections of its asymmetric part $P= R\backslash R^{-1}$ are open.  
   We shall also denote $R$ by $\preceq$, as is standard especially in the economics  literature.

The descriptive adjectives pertaining to a relation are presented in a tabular form for the  convenience of the reader in  Table \ref{tbl: relation} below. 
  
 \medskip
 
\begin{table}[ht]
\begin{center}
\begin{tabular}{lll} 
 & Set-theoretic notation & Relational notation\\ 
\hline  \noalign{\vskip 1mm}     
{\it reflexive} & $\Delta\subset R$ & $x\preceq x$ $\forall x\in X$\\ 
{\it complete}& $X\times X=R\cup R^{-1}$  & $x\preceq y$ or $y\preceq x$ $\forall x,y\in X$\\ 
{\it symmetric}&  $R=R^{-1}$  & $x\preceq y$ implies $y\preceq x$  $\forall x,y\in X$\\ 
{\it asymmetric} &  $R\cap R^{-1}=\emptyset$   & $x\preceq y$ implies $y\not\preceq x$ $\forall x,y\in X$\\ 
{\it anti-symmetric} &  $R\cap R^{-1}\subset\Delta$   & $x\preceq y\preceq x$ implies $x=y$\\ 
{\it non-trivial} & $R\backslash R^{-1}\neq\emptyset$ &  $\exists x,y\in X$ such that $x\preceq y$ and $y\npreceq x$ \\ 
{\it transitive} & $R^{-1}(y)\times R(y)\subset R$  $\forall y\in X$  & $x\preceq y\preceq z$ implies $x\preceq z$ $\forall x,y,z\in X$\\ 
{\it negatively transitive} & $R^c$ is transitive &$x\not\preceq y\not\preceq z$ implies $x\not\preceq z$  $\forall x,y,z\in X$ \\
\hline
\end{tabular}
\end{center}
\vspace{-10pt}

\caption{Properties of Binary Relations}
\lb{tbl: relation}
\end{table}  
  
\vspace{-6pt}


\noindent As is conventional, we denote the {\it symmetric part} of the binary relation  $R$ by  $I = R\cap R^{-1}$ and its {\it asymmetric part} by $P = R\backslash R^{-1}.$    In terms of  \cites{se69}  felicitous   notation, let  $P\! I$ be the containment $P^{-1}(x)\times I(x)\subset P$ for all $x\in X,$  $I\! P$ the containment $I^{-1}(x)\times P(x)\subset P$ for all $x\in X.$  Note, for example, that   $P\! I$ amounts to the requirement that $y\prec x\sim z$ implies $y\prec z$ and $I\! P$ that $y\sim x\prec z$ implies $y\prec z,$ where $\sim$ denotes the symmetric part of $\preceq$ and $\prec$ its asymmetric part. 
 We shall also use the abbreviations $T, \; P\! P, \; II$ and $N\! P,$ where the first three refer to the transitivity of $R, P$ and $I$, respectively, and $N\! P$ the negative transitivity of $P.$  

%
%

 \df  A binary relation is said to be {\it semi-transitive} if $P\! I$ and $I\! P$ hold.  \dff

\df
A topological space $X$ is said to be {\it connected} if it is not the union of two non-empty, disjoint open sets. Equivalently,  $X$ is connected if the only subsets of $X$ which are both open and closed are $\emptyset$ and $X.$ The space $X$ is {\it disconnected} if it is not connected. A subset of $X$ is connected if it is connected as a subspace. 
\lb{df: connected}
\dff

All this is routine so far. We now break new ground by considering the concepts of $k$-connectedness of a set and $k$-nontriviality of a binary relation on that set.\fn{We do not mean to overemphasize the novelty: a cursory reference to \cite{wi49}, or to \cite{du66}, shows that  spaces with finitely many components are  well-known and  studied extensively. The reader should beware, however, that  $k$-connectedness  is used in algebraic topology with a  totally different meaning.}

\df
A component of a topological space is a maximal connected set in the space; that is, a connected subset which is not properly contained in any connected subset.\fn{There is the weaker concept of a {\it quasi-component} of a topological space. In a compact Hausdorff space, the two are  identical; see \cite{wi49}. Even though many economic settings assume the Hausdorff separation axiom, compactness may fail, like the consumption set of a consumer in a Walrasian economy. Hence, it may be of interest to see if the results in this paper generalize to quasi-components. \label{fn:quasi}} For any natural number $k,$ a topological space is $k$-connected if it has at most $k$ components. 
\lb{df: comp}
\dff
\noindent   The concept of  $k$-connectedness provides a quantitative measure of the degree of disconnectedness of a topological space. Note also that it is an adjective for space, and not for the topology it is endowed with. It is also clear that $1$-connectedness is equivalent to connectedness and that any $k$-connected space is $l$-connected for all $l\geq k.$  
\smallskip

 Next, we present a generalization of non-triviality. 
\vspace{-3pt}

\df
A binary relation $R$ on a topological space $X$ is {\it componentwise non-trivial} if 
\ben[{\nf (i)},  topsep=3pt]
\setlength{\itemsep}{-1pt} 
\ml for any component $C$ of $X,$ there exists $x,y\in C$ such that $(x,y)\in P,$ 
\ml for any distinct components $C, C'$ of $X,$ there exist $x\in C, y\in C'$ such that $(x,y)\in R\cup R^{-1}.$ 
\een
\lb{df: cn}
\dff
\vspace{-3pt}

\noindent It is easy to see that the concept requires strict comparability within the components and weak comparability across the components.  We now give it a quantitative cast by presenting a formal definition of $k$-nontriviality. 

\df
Let $X$ be a topological space and $\{C_l\}_{l\in L},$ $L$ an arbitrary index set, denote its components. For any finite $k\leq |L|,$ we say that  a binary relation $R$ on $X$ is $k$-non-trivial if there exist $1\leq m_1<\cdots<m_k\leq k$ and $1\leq n_1<\cdots<n_k\leq k$ such that  for all $i,j\leq k$, $i<j$,
\ben[{\nf (a)},  topsep=3pt]
\setlength{\itemsep}{-1pt} 
\ml  there exists $(x,y)\in C_{m_i}\times C_{n_i}$ such that $(x,y)\in P\cup P^{-1},$
\ml  there exists $(x,y)\in (C_{m_i}\times C_{n_j})\cup (C_{m_j}\times C_{n_i})$ such that $(x,y)\in R\cup R^{-1}.$
\een
\lb{df: knontrivial}
\dff
\vspace{-3pt}

\nt  First, the elementary observation that in any  space,  non-triviality and  $1$-non-triviality of a binary relation are equivalent. Furthermore, for  $k$-connected spaces, for all $i\leq k,$  $k$-non-triviality of a binary relation can be qualitatively conceived of as being componentwise  non-trivial since $m_i=n_i=i.$ Nevertheless, the concept is not straightforward, and it is worthwhile to discuss it in the context of simple examples: we  illustrate  {\it 1-, 2-} and {\it 3-non-triviality} in the context of a space with three components.  


\begin{figure}[ht]
\begin{center}
 \includegraphics[width=6.4in, height=2.1in]{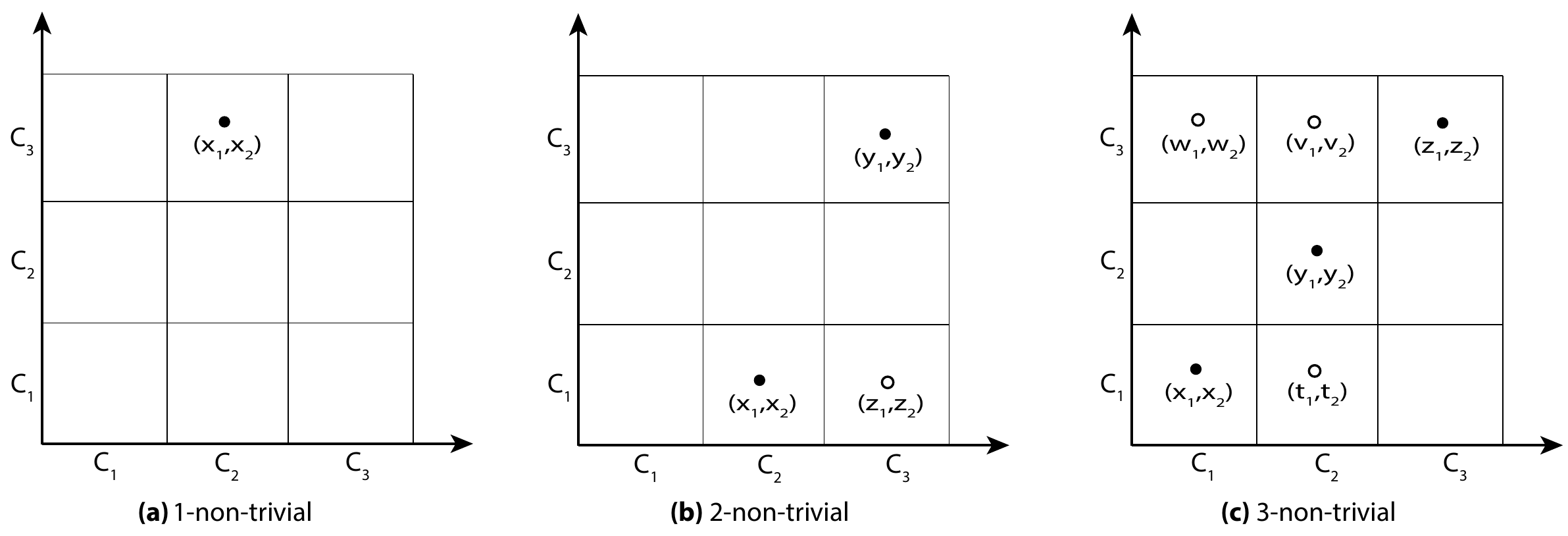}
\end{center}  
\vspace{-20pt}
    
%
\caption{$k$-nontriviality}    

\lb{fig: nt}
\end{figure}
\vspace{-5pt}

\exm {\nf 
In Figure  \ref{fig: nt}, $X=C_1\cup C_2\cup C_3$ is the union of three non-degenerate open intervals in the real line which is endowed with its standard  topology. It is clear that $C_1, C_2, C_3$ are the components of $X$. Panels (a), (b) and (c) of Figure illustrate three distinct binary relations $R_a, R_b, R_c$ defined on $X$, with  points labeled with a filled circle illustrating  condition (a) of Definition \ref{df: knontrivial},  and the ones with an empty circle illustrating  condition (b).  Hence,  $R_a=P_a, R_b=P_b$ and $R_c=P_c$. We now turn the detailed explanation of each relation illustrated in the panels. 


 We first show that the relation $R_a$ is 1-non-trivial. In Definition \ref{df: knontrivial},  set $k=1, m_1=2, n_1=3$. As illustrated, $(x_1, x_2)\in C_{m_1}\times C_{n_1}$. Since $(x_1, x_2)\in  R_a$ and $(x_2,x_1)\notin R_a$, therefore $(x_1, x_2)\in P_a$. Therefore, condition (a) of the definition is satisfied. Condition (b) does not have any  bite since  there are no distinct $i,j\leq k=1$.  
 
 Second, we show that the relation $R_b$ is 2-non-trivial. In this example, it is clear that $R_b=P_b$. Set $k=2$, $m_1=2, m_2=3$ and $n_1=1, n_2=3$. Since each of $C_{m_1}\times C_{n_1}$ and $C_{m_2}\times C_{n_2}$  contains a pair of strictly comparable  alternatives, $(x_1, x_2)$ and $(y_1, y_2)$ respectively,  $R_b$ satisfies condition (a) of the Definition  \ref{df: knontrivial}.   The only $i,j$ that satisfy $i<j\leq 2$ are $i=1, j=2$. It follows from $(z_1,z_2)\in P_b\cap C_{m_2}\times C_{n_1}$ that condition (b) also holds. Hence, $R_b$ is 2-non-trivial.
 
 Finally, we  show that the relation $R_c$ is 3-non-trivial. In this example, it is clear that $R_c=P_c$. Set $k=3$. In this case, there exists only one ordering that satisfies  $1\leq l_1<l_2<l_3\leq 3$ which is  $1<2<3$. Hence $m_i=n_i=i$ for $i=1,2,3$ in the definition above.  Since for each $i$, each $C_i$ contains a pair of strictly comparable points (one of $(x_1, x_2), (y_1, y_2)$ and $(z_1, z_2)$),  condition (a) of 
Definition  \ref{df: knontrivial}   holds. For $i=1<2=j$, $(t_1, t_2)\in C_{m_j}, C_{n_i}$, for $i=1<3=j$, $(w_1, w_2)\in C_{m_i}, C_{n_j}$ and for $i=2<3=j$, $(v_1, v_2)\in C_{m_i}, C_{n_j}$. Therefore,  condition (b) also holds. 
Example \ref{exm: nontrivial} is complete.    \qed
\lb{exm: nontrivial}
}\exmm


\section{The Theory} \lb{sec: main} 
In this section, we present the six basic results of this paper along with  two supplementary propositions. The first three subsections present equivalence results characterizing connectedness and its natural generalizations; while the fourth is devoted to two sufficiency theorems. These two theorems set Sen's  topologically-free results in a topological register, and elaborate Sonnenschein's invocation of the Phragmen-Brouwer (topological) property, all in the setting of incomplete preferences. These six theorems all have constitutive implications for what we are referring  to as the ES research program.\fn{In terms of a reader's guide, and especially for a first reading, one may read these theorems for the case $k=1,$ and instead of the full equivalence, even limit oneself to the {\it forward} direction  as in the foundational papers.  \label{fn:rg1}}

\subsection{A General Result on $k$-Connectedness} \lb{sec: con}

We now present our  first  equivalence theorem, deferring to the next subsection its relationship to the antecedent literature.

%

\thm
For any natural number $k,$  the following statements are equivalent for a  $k$-non-trivial and continuous binary relation defined on any  topological space with at least $k$-components.   
\ben[{\nf (a)},  topsep=3pt]
\setlength{\itemsep}{1pt} 
\ml The space  is $k$-connected. \lb{it: kc} 
\ml Any  transitive relation  is complete. \lb{it: ksc}
\ml Any  anti-symmetric relation  is complete. \lb{it: keg}
\ml Any relation whose symmetric part is transitive with connected sections,  is complete. \lb{it: kes}
\ml Any  semi-transitive relation with  transitive symmetric part  is complete. \lb{it: kess}
\een
\lb{thm: kc}
\thmm

%

Even though Theorem \ref{thm: kc} achieves a symbol-free and clear expression, a further discussion of it in terms of the vernacular used in the introduction may be worthwhile.    But before that, note that the theorem works with a class of topologies, and that  statement (a),  together with the specification that the space has at least $k$-components,  implies that the space has {\it exactly} $k$ components. Moving on to the other assertions,   under the topological specifications of connectedness and continuity,  condition~\ref{it: sc} asserts  the impossibility of a consistent agent being indecisive, and condition \ref{it: eg} insists that she, when  never  indifferent between  distinct items,  must be  decisive irrespective of being consistent or not. The third assertion drops consistency and adds extreme decisiveness regarding comparable  items. Indeed, one can push  the latter claim  a little further.   Even when she can ``choose and not only pick"\fn{We are indebted to \cite{um77}  for this terminological distinction; also taken up in \cite{se93}, 
 and under the heading of \lq\lq the idea of internal consistency of choice," in \citet[pp. 309-312]{se17}.}  which is to say, when she has a fine enough  taste  to discriminate between distinct items and to be possibly indifferent among them (has possibly non-singleton transitive indifference classes),   she must be consistent if these  classes are either connected (condition~\ref{it: es}), or can be compared in the sense of being semi-transitive  (condition~\ref{it: ess}).  As Figure~\ref{fig: inclusion} makes clear, there is a relationship of inclusion\fn{We shall return to this below in Proposition \ref{thm: transitivity} and in Theorems \ref{thm: ttransitivity} and \ref{thm: pbtransitivity}.}  in terms of conditions  \ref{it: sc} to  \ref{it: ess}  of Theorem \ref{thm: kc};  and so, at least as far as the {\it forward} direction from condition~\ref{it: c} is concerned, the statement  \ref{it: c}$\Rightarrow$\ref{it: ess} is all that needs to be shown. However, the very generality of condition~\ref{it: ess} goes towards moving away from it when we consider the {\it backward} direction.

\begin{figure}[ht]
\begin{center}
 \includegraphics[width=4in, height=1.8in]{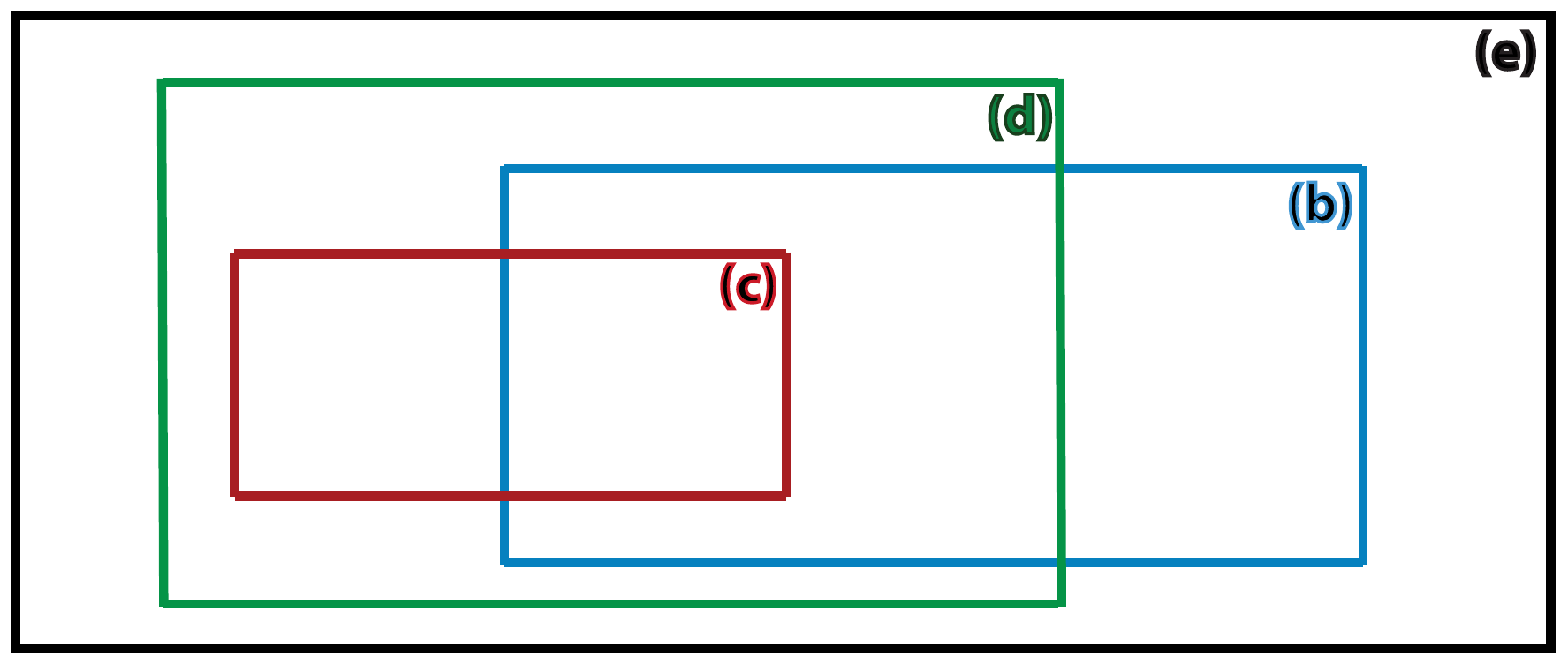}
\end{center}  
\vspace{-7pt}
    
\footnotesize{The sets {\bf (b)}--{\bf (e)} corresponds to the sets of binary relations which satisfy the assumptions of assertions {(b)}--{(e)} of Theorem \ref{thm: kc}, respectively. In particular, the set {\bf (e)} denotes the set of all $k$-non-trivial, semi-transitive and continuous binary relations $R$ whose symmetric part $I$ is transitive, {\bf (d)} is the subset of (e) such that the sections of $I$ are connected, 
     {\bf (c)} is the subset of {(e)} such that $R$ is anti-symmetric and  {\bf (b)} is the subset of {(e)} such that $R$ is transitive.} 
   
\vspace{-5pt}   
    
\caption{The inclusion relationship between assertions  \ref{it: sc}--\ref{it: ess} of Theorem \ref{thm: kc}}    

\lb{fig: inclusion}
\end{figure}



\subsection{Specializations:  Connectedness and 2-Connectedness} \lb{sec: c2c}



For the special cases of connected and 2-connected spaces,  we have more interesting equivalence results. In terms of the forward direction, we not only obtain completeness but also transitivity for free!


\thm 
For any natural number $k\leq 2,$  the following statements are equivalent for a  $k$-non-trivial and continuous binary relation defined on any  topological space with at least $k$-components.     
\ben[{\nf (a)},  topsep=3pt]
\setlength{\itemsep}{1pt} 
\ml The space is $k$-connected. \lb{it: c}
\ml Any transitive relation  is complete.  \lb{it: sc}
\ml Any  anti-symmetric  relation is complete and transitive. \lb{it: eg}
 \ml Any relation whose  symmetric part is  transitive with  connected sections,  is complete and transitive. \lb{it: es}
\ml Any  semi-transitive relation with a transitive symmetric part  is complete and transitive. \lb{it: ess}
\een
\lb{thm: c}
\thmm


\noindent Thus, once we specialize to connectedness or to 2-connectedness, we can strengthen Theorem 1 to obtain consistency {\it and} decisiveness instead of only decisiveness. 
 As already emphasized in the introduction, a simple  gathering of the foundational results has led to an equivalence theorem,  and thereby also to a characterization of {connectedness} and {2-connectedness} of the choice set. In terms of the relationship of Theorem 2 to the antecedent literature, it takes a piecemeal treatment  into a mutually reinforcing one, and thereby in indissolubly  connecting the behavioral and mathematical registers, testifies to the analytical depth of the ES research program.\fn{We may point out here in anticipation that our joint treatment of these foundational results  has also allowed  us to provide alternative proofs of the results of Eilenberg and Sonnenschein. Our method of proof is inspired by Schmeidler, and is totally different compared to theirs; see this observation formally made in the paragraph preceding the proof of this theorem, and in the remark following it.} 

Next, moving to a blow-by-blow account of a comparison with the theorems of  Eilenberg, Sonnenschein and Schmeidler, it is perhaps easiest to begin with the concept of {\it semi-transitivity,} pioneered, though not named as such,\fn{Note that the usage here differs from that of Houthakker's concept; see  \citet[pp. 94, 295-298, 299]{se17}. \label{fn:semi}} by \citet[p. 232]{ra63}, and formally presented as Definition 1 above.  He used it to present what he saw to be a  \lq\lq remarkable result, due, in essence to Eilenberg," and referred to it as a condition under which the choice set 

\bqu  may be decomposed into indifference classes and that these classes
may be compared as more or less preferable. No strict transitivity assumption is made,
although two indifference classes are never allowed to be indifferent to each other. \equ 

\noindent In  the specialization of Theorem \ref{thm: c} to $k=1$, note that 
 \vspace{1pt}

\ben[{\nf (i)},  topsep=3pt]
\setlength{\itemsep}{1pt} 
\ml   \ref{it: c}$\Rightarrow$\ref{it: ess}  drops the completeness and anti-symmetry assumptions of  \citet[2.1]{ei41} and the completeness assumption of \citet[Theorem 3]{so65},   weakens the transitivity assumption of \citet[Theorem]{sc71};  
\ml   \ref{it: c}$\Rightarrow$\ref{it: es} drops the completeness and anti-symmetry assumptions of  \citet[2.1]{ei41} and the completeness assumption of  \citet[Theorem 4]{so65};   
\ml   \ref{it: c}$\Rightarrow$\ref{it: eg} drops the completeness assumption of \citet[2.1]{ei41}; 
\ml \ref{it: c}$\Rightarrow$\ref{it: sc} is due to \citet[Theorem]{sc71}. 
 \een
 \vspace{4pt}
 
All this raises a larger point that   some of the assumptions that we make salient are {\it hidden} in the statements of the theorems of Eilenberg, Sonnenschein and Schmeidler. For example, Schmeidler's transitivity assumption already implies that the relation is semi-transitive and its symmetric part  is transitive. As another example, Sonnenschein does not assume that the symmetric part of a relation is transitive, but under the completeness assumption, this already follows from the relation itself being semi-transitive.\fn{On this, see Figure~\ref{fig: inclusion} and Proposition \ref{thm: transitivity}. 
 These relationships are partially available  in \citet[Theorem 3]{so65}, \citet[Theorem 2]{lo67} and \citet[Theorem I]{se69}.}    The reader should also note that the  theorem of Eilenberg, as well as that of Sonnenschein, does not assume non-triviality; but cases where non-triviality does not hold,  yield trivialities in themselves.\fn{We return to this in Section 4.1. \lb{fn:triviality}}    If  the preference relation is trivial in Eilenberg's result, then the choice space can consist of at most one element. And in Sonnenschein's result, it  requires all elements to be indifferent to each other. In both cases, therefore, transitivity of the preference relation is equivalent to the transitivity of the indifference relation which is already assumed.
  This issue of  {\it hidden} assumptions goes beyond these foundational papers,\fn{To be sure, the discussion in this paragraph assumes $k=1;$ the foundational literature is silent on higher values of the natural number $k.$}  and in our consideration of other work,  we shall return to this theme  below in Section \ref{sec: app}.


Two final remarks concerning Theorem \ref{thm: c}. First, assertions \ref{it: eg}--\ref{it: ess} in Theorem \ref{thm: c} reproduce assertions \ref{it: keg}--\ref{it: kess}  in Theorem \ref{thm: kc}, but with the strengthened form that substitutes completeness {\it and} transitivity for completeness. The reader should note that this strengthening does not necessarily hold for $k>2.$  The following Example settles this issue.

\exm {\nf Let $X=(0,1)\cup (1,2)\cup (2,3)$ and the topology is the Euclidean metric. Then, $k=3.$ Now let $R$ be an anti-symmetric binary relation defined as follows: $(x,y)\in R$ if $x,y\in C_k, x\leq y,$ if $x\in (0,1)$ and $y\in (1,2),$ if $x\in (1,2)$ and $y\in (2,3),$ and if $x\in (2,3)$ and $y\in (0,1).$   It is clear that $R$ is complete and has closed sections. However, it is non-transitive.    \qed
 %
\lb{exm: 2k}
}\exmm

\nt Our second remark concerns assertion \ref{it: sc}. Even though  it already assumes transitivity and is included in Theorem \ref{thm: kc}, it does not mean it is redundant because of the basic issue of equivalence. 
\vspace{-8pt}

 %
In Theorems 1 and 2,   the implication \ref{it: c} $\Rightarrow$ \ref{it: sc} is a literal rendering of Schmeidler's claim of completeness, in the case $k=1.$  Moreover, the Eilenberg-Sonnenschein claim of transitivity is ignored in this first result  and bundled with completeness in the second. 
   The question then arises as to whether there is an equivalence claim that is focused only on transitivity. This is to ask whether in assertions (c)--(e) in  Theorem 2, completeness can  be  shifted  from the  conclusion to the hypothesis. The answer is negative.   
   In order to see this, let $X=\{0,1\}$ be endowed with a discrete topology. Then it is clear that $X$ is disconnected and every binary relation on $X$ has both closed and open sections, and is transitive. Therefore, even though connectedness of the space is a sufficient condition for the results of Eilenberg-Sonnenschein, it is not a necessary condition. However, in our simple example, the space is 2-connected, and this gives a hint that a version of Theorem \ref{thm: c}  may be true for 2-connected spaces when the completeness of the relation is correspondingly shifted. It is indeed so, but involves a subtlety to which we turn next. 
   
 As already emphasized in the introduction, Eilenberg's  paper pioneered the question of the representation of a preferences relation on a set in terms of a real-valued function on the same set, but it also investigated the question of conditions under which a  binary relation with attractive natural properties exists on the set.\fn{We have already emphasized the first question in Footnote~\ref{fn:ei} and in   Footnote~\ref{fn:rep}, but what is being emphasized  here is  the existence of the relation rather than a function, a question that is taken up in the social choice literature; see \cite{ku18} for an exploration of this question and an elaboration of this connection.   At any rate, as we shall see, it is essential to the result we present below.   \label{fn:sc}}  Introducing   a notion of an {\it ordered} space as one that admits an   anti-symmetric, complete, transitive and continuous relation, he presented the following result. 

\bqu {\it A connected topological space $X$  containing  at least two elements can be ordered 
  if and only if $P(X)$ is disconnected, where $P(X)$ consists of $(x,y)\in X\times X$ such that $x\neq y.$ }
\equ

\nt   In the  move  to our next result, we introduce the following definition which drops the transitivity requirement in  \cites{ei41} characterization of an ordered space.\fn{The reader is again warned about the lack of a uniform terminology in the literature; see \citet[p. 54]{se17}.} 

\df
A topological space is {\it quasi-ordered} if there exists an anti-symmetric, complete and continuous binary relation on it. 
 \lb{df: ordered}
\dff

We are now  ready to state the third equivalence result of the paper.  

\thm
The following statements are equivalent for a complete and continuous  binary relation on any quasi-ordered topological space. 
\ben[{\nf (a)},  topsep=3pt]
\setlength{\itemsep}{1pt} 
\ml The  space  is 2-connected. \lb{it: 2cb}   
\ml Any anti-symmetric relation is transitive.    \lb{it: eil}
\ml Any  relation whose symmetric part is transitive with connected sections, is transitive. \lb{it: son2}
\ml Any semi-transitive relation is transitive.  \lb{it: son}
\een
\lb{thm: 2cb}
\thmm

The subtlety referred to above  involves the restriction of the result to a quasi-ordered topological space. To begin with,  note at the outset that the restriction is required only for the  {\it backward} direction, and that to ensure that  the assertions guaranteeing  
2-connectedness are not vacuous, that there indeed exist binary relations satisfying the properties required  of them. Eilenberg's theorem guarantees their existence in an ordered space, and {\it a fortiori}, in a quasi-ordered space.  Note this is not an issue in Theorems 1 and 2 since we can construct a $k$-non-trivial continuous binary relation without any assumption on the topological space.\fn{This construction is illustrated in the backward direction of the proofs of Theorems 1 and 2; see the penultimate paragraph of the proof of Theorem 1 presented below. As regards Theorem 2, the cases $k=1$ and $k=2$ are considered separately; see the last paragraph of the proof in each case.}

Proposition \ref{thm: discrete} in Section \ref{sec: app} implies that the topology on a finite topological space with an anti-symmetric, complete and continuous binary relation (quasi-order) has to be discrete. This shows that there does not exist a quasi-order on any finite topological space which is not discrete. We next illustrate an example of such a topological space which has three components. On this space there does not exist a quasi-order, hence it is vacuously true that any quasi-order is transitive. But the space is not 2-connected.  

\exm {\nf Let $X=\{x,y,z,w\}$ and the collection $\{\emptyset, \{x\}, \{y\}, \{z,w\}\}$ is a basis for the topology defined on $X$. For any complete and anti-symmetric binary relation $R$ on $X$,  either $(z,w)\in P$, or $(w,z)\in P$, where $P$ is the asymmetric part of $R$. Then, either $w\in P(z)$, or $w\in P^{-1}(z)$. Note that $z\notin P(z)\cup P^{-1}(z)$ by definition. Then, either $P(z)$, or $P^{-1}(z)$ contains $w$ but  excludes $z$. Since continuity of $R$ requires $P$ has open sections, any complete and anti-symmetric relation on $X$ cannot be continuous.   \qed
\lb{exm: quasiorder}
}\exmm

\nt For an example of a general topological space, we know by Proposition \ref{thm: discrete} that there does not exist a quasi-order on any non-Hausdorff topological space. Such examples are analogous to the one above.   

 And while we are on the asymmetry between Theorem 3 and the the preceding Theorems 1 and 2, let us also note that for a symmetrical treatment, we can simply include assertion (b) in the theorem  by specifying attention to 2-non-trivial binary relations, and taking the requirement of completeness down to the other assertions.\fn{We do not do this because Theorem 3 is primarily motivated by the transitivity claim of Eilenberg-Sonnenschein.}  
In Theorem \ref{thm: 2cb},  the statements \ref{it: 2cb} $\Rightarrow$  \ref{it: eil}, \ref{it: son2}, \ref{it: son} are generalizations of \citet[Theorem]{sc71}, \citet[2.1]{ei41} and \citet[Theorems 4 and 3]{so65} to 2-connected spaces, respectively.\fn{We present here a slightly stronger version of Sonnenschein's theorems in order to underscore the comparison. He assumes the connectedness of $X\backslash I$ instead of connectedness of $X.$ It is easy to prove a version of our results under this assumption. A result similar to 
 Sonnenschein's theorem is provided by \citet[Lemma]{ra63} under a stronger connectedness assumption which requires the upper sections of the weak preference relation to be connected.} For the reader's convenience,   Table \ref{tbl: comparison} summarizes the statements of Theorems 1-- 2 as well as those of Eilenberg, Sonnenschein and Schmeidler which are listed in Theorem 3.

\begin{table}[ht]
{
\begin{center}
\begin{tabular}{lcccccccccccc}
\hline \hline \\[-2ex]
   & Eil & Son & Sch  &  \!\! T\ref{thm: kc}\ref{it: ksc} & \!\! T\ref{thm: kc}\ref{it: keg} & \!\!\! T\ref{thm: kc}\ref{it: kes}  & \!\!\!  T\ref{thm: kc}\ref{it: kess} & \!\!\! T\ref{thm: c}\ref{it: eg} & \!\!\! T\ref{thm: c}\ref{it: es}  & \!\!\! T\ref{thm: c}\ref{it: ess} \\ 
\hline \\[-2ex] 
 $X:$ (2-)connected & \checkmark & \checkmark & \checkmark  & $\times$  & $\times$&  $\times$& $\times$ & \checkmark & \checkmark & \checkmark \\ 
 ~~~~ $k$-connected & \checkmark & \checkmark & \checkmark & \checkmark  & \checkmark  & \checkmark  & \checkmark  & \checkmark & \checkmark & \checkmark \\ 
 $R:$ complete & \checkmark & \checkmark &  $\times$& $\times$ & $\times$ & $\times$ & $\times$  & $\times$ & $\times$ & $\times$ \\ 
~~~~ transitive & $\times$& $\times$ &\checkmark &\checkmark &  $\times$ & $\times$ & $\times$ &  $\times$ & $\times$ & $\times$  \\ 
~~~~ semi-transitive & \checkmark & \checkmark & \checkmark &  \checkmark & \checkmark & $\times$ & \checkmark &  \checkmark & $\times$ & \checkmark \\ 
~~~~ anti-symmetric & \checkmark & $\times$  &  $\times$ & $\times$  & \checkmark & $\times$   & $\times$  &  \checkmark & $\times$ & $\times$ \\ 
~~~~ continuous & \checkmark & \checkmark  & \checkmark & \checkmark & \checkmark & \checkmark & \checkmark  &\checkmark & \checkmark & \checkmark \\
~~~~ (2-)non-trivial$^*$ & \checkmark & \checkmark  
 & \checkmark & \checkmark & \checkmark & \checkmark & \checkmark &  \checkmark & \checkmark & \checkmark \\
 ~~~~ $k$-non-trivial  & $\times$ & $\times$ & $\times$  & \checkmark & \checkmark & \checkmark & \checkmark &  $\times$  & $\times$ & $\times$ \\
$I:$  transitive & \checkmark & \checkmark & \checkmark & \checkmark &  \checkmark & \checkmark & \checkmark  & \checkmark & \checkmark &  \checkmark \\
~~~ connected sections & \checkmark & $\times$ 
 & $\times$ & $\times$ & $\times$ & \checkmark & $\times$  & $\times$ & \checkmark & $\times$  \\
[1ex]
\hline
\end{tabular}
\vspace{5pt}

 \raggedright \footnotesize{In this table, if one property directly follows from the other assumed properties, then we mark it as \checkmark. For example, Eilenberg's theorem assume the preference relation is complete and anti-symmetric, which directly imply semi-transitivity of the preference relation along with the transitivity of its symmetric part. Hence we mark that these two additional properties are also assumed. 
 
 * Even though theorems of  Eilenberg and Sonnenschein do not assume (2)-non-triviality, without this assumption their results become triviality, see Section \ref{sec: c2c} for details.}

\vspace{-5pt}
  
\caption{Comparison of the Results}
\lb{tbl: comparison}
\end{center}
}
\end{table}


\cite{lo67} claims that Sonnenschein's theorems are not really theorems of topology, but  theorems of set theory that can be proved by set-theoretic considerations alone. In his reply to Lorimer, \citet{so67} provides a version of assertion  \ref{it: 2cb} $\Leftrightarrow$ \ref{it: son} of Theorem \ref{thm: 2cb} for subsets of the real line in order to highlight the necessity of connectedness. Sonnenschein also adds that Lorimer's conditions ``would have been very unnatural and generally of little use in economic applications."  The point is that Eilenberg and Sonnenschein use continuity and connectedness assumptions at a crucial step in their results, and  Lorimer simply assumes that that step holds in order to eliminate the topological register and thereby limit himself \lq\lq solely" to a set-theoretic one.\fn{It should be noted, however,  that Lorimer also shows that these conditions are not only sufficient for transitivity of a complete relation, but also necessary, and hence there is something extra and of consequent use  in his paper.} Along this line of investigation,  \citet[8.1]{ei41} and \citet[Theorem 5]{wa54} provided results on the necessity of topological connectedness by using order-structure properties. However, their results are partial.

Next, to allow the reader some breathing space, and to compare what we have achieved so far relative to the three foundational results laid out in the introduction, we 
 summarize the results obtained from the  forward direction of the three theorems: a sufficient condition for completeness and transitivity of a binary relation defined on an {\it arbitrary} topological space.  The essential idea has little need of the index $k$ for its articulation, and hence attests to the fact that the forward direction is true for any topological space. 
 %

\prp
 Every componentwise non-trivial, semi-transitive and continuous binary relation on  a topological space, whose symmetric part is transitive, is complete. If the topological space has at most two components, then the binary relation is also transitive. 
\lb{thm: cw}
\prpp
 
\noindent This proposition shows that if there are comparable alternatives within and across components, then all alternatives are comparable.  The proof of the necessity of completeness follows from the proof of Theorem \ref{thm: kc} since its proof requires only componentwise non-triviality and does not hinge on the finiteness of the number of components. The necessity of transitivity follows from Theorem \ref{thm: c}. Moreover, Theorem \ref{thm: c}\ref{it: son} implies that semi-transitivity can be substituted with  further topological assumptions on the binary relation; see Section 3.4 below for details, in particular Theorems \ref{thm: ttransitivity} and \ref{thm: pbtransitivity}.


\subsection{A Return to Connectedness} 

A natural question to ask at this stage is how Theorem \ref{thm: c} can be further sharpened if we limit ourselves to the setting original to the foundational  papers. This is to say, to $1$-connectedness of the topological space. This, by necessity, brings us to current work, and allows a re-framing and a generalization of it.\fn{We are primarily motivated by forging connections to current work; there is little doubt that results of several of the papers that we connect to can be generalized to the setting of $k$-connected sets for any natural number $k$, and thereby to reformulate the theorem presented below along the lines of Theorem 1.}

Towards this end,  we begin with a  stronger non-triviality concept.

\df
A binary relation $R$ on a set $X$ is called {\it strongly non-trivial} if there exists $x\in X$ such that  $P(x)\neq\emptyset,$ and  $R(x')\cap R(y')\neq \emptyset$ for all $x',y'\in P(x).$
\lb{df: snt}
\dff

\nt In terms of relational notation, strong non-triviality amounts to the requirement that (i)  there exist $x, y\in X$ with $x\prec y$, and that (ii)  for all $x',y'\in X$ with $x\prec x'$ and $x\prec y'$, there exists $z\in X$ such that $x'\prec z$ and $y'\prec z.$

 It is not surprising that weakening the continuity assumptions in the hypothesis of  Theorems 1 to 3 falsifies    their conclusions. In order to elaborate on this, we consider  the concept of {\it fragility}  due to \cite{ge13}, and supplement it by a parallel concept of   {\it flimsiness:}  both are useful for the  analysis of the structure of the preferences under weaker continuity assumptions.  The first  assumes that every neighborhood of some strictly comparable alternatives of a non-trivial  binary relation on the choice space contains non-comparable alternatives,  \citet[p. 161]{ge13} motivates his concept as follows: 

\bqu From a normative point of view, fragility of a preference relation is an undesirable property, both when this preference relation is that of an individual and also when it represents the preferences over social alternatives. 
Indeed, one would expect that when decision makers express strict preference for one alternative over another, marginal changes in these two alternatives should not result in them becoming incomparable. If they do, then doubt should perhaps be cast on the validity of the strict-preference comparison between the original alternatives. Finally, introspection and casual empiricism do not seem favorable for the property's descriptive accuracy either.
\equ

\df
We call a binary relation $R$ on a topological space $X$ fragile if there exist $x,y\in X$ such that 
 {\nf (i)}  $(x,y)\in P,$ and that {\nf (ii)}  any open neighborhood of $(x,y)$ contains $(x',y')$ such that $(x',y')\notin R\cup R^{-1}.$  \lb{df: fragile}
\dff 


\nt In terms of relational notation, fragility amounts to the requirement that
(i)    there exists $x, y\in X$ with $x\prec y$, and that (ii) 
any open neighborhood of $(x,y)$ contains $(x',y')$ such that  $x'\npreceq y'$ and $y\npreceq x'.$
\smallskip

 Next, we provide a definition which complements fragility. 
\df
A binary relation $R$ on  a topological space $X$ is flimsy if there exist $x,y\in X$  with $(x,y)\notin R\cup R^{-1}$ such that every open neighborhood of $(x,y)$ contains $(x',y')\in  R\cup R^{-1}$.
\lb{df: sensitive}
\dff

\nt In terms of relational notation, flimsiness amounts to the requirement that
  there exists $x, y\in X$ with $x\npreceq y$ and $y\npreceq x$  such that  every open neighborhood of $(x,y)$ contains $(x',y')$ with $x'\preceq y'$ or $y'\preceq x'$.
\smallskip

 \citet[Corollary 3]{ge13}  showed that every incomplete, non-trivial and transitive binary relation with closed sections defined on a connected topological space, is fragile. Therefore, dropping one of the continuity assumption of Schmeidler's theorem yields an undesirable case of incompleteness from the normative point of view. We show that his result is equivalent to topological connectedness of the space.  And we supplement his result by other affiliated concepts; but before turning to them, we  recall for the reader three ways of  taking an asymmetric relation and associating  its reflexive hull with it.     

\df  For any  asymmetric binary relation $P$  on a topological space $X,$  define its reflexive hull as $R=\left(P^c\right)^{-1}=\{(x,y)~|~(y,x)\notin P\}.$ 
 Moreover, define the lower covering relation $R_\ell$ of $R$ and its upper covering\fn{These ``covering'' type of derived relations have important applications and implications in decision theory as well as in social choice theory. In particular,  they have been used for the numerical representation of incomplete preferences; see for example \citet{ch87} for representation of interval orders, and \citet{pe70} and \cite{gk13} for expected utility representation of incomplete preferences.  In social choice theory, the ``covering relation'' has been used since mid-twentieth century; see \cite{du13} for a comprehensive survey on covering relation. \lb{fn: covering}} 
 relation $R_u$  as follows: 
\vspace{-5pt}
\[
\begin{array}{lcl}
R_\ell=\{(x,y)~|~ R^{-1}(x)\subset R^{-1}(y)\}  &\mbox{and} & R_u=\{(x,y)~|~ R(y)\subset R(x)\},\\

\end{array}
\]
\lb{df: hull}
\dff
\vspace{-25pt}
\nt We call the pair $\left(R_\ell, R_u\right)$ the {\it covering relations} of $R$. In terms of relational notation, the reflexive hull of an asymmetric relation $\prec$ is defined as follows: $x\preceq y$ if and only if $x\nprec y.$ The lower covering relation $\preceq_\ell$ of $\preceq$ and its upper covering relation $\preceq_u$ are defined as follows: $x\preceq_\ell y$ if and only if $z\preceq x$ implies $z\preceq y$, and $x\preceq_u y$ if and only if $y\preceq z$ implies $x\preceq z$. 

\medskip

 Finally, we present two additional concepts due to \cite{ch87}, the second being his strengthening of the standard notion of a separable relation. 

\df  Let $P$ be an asymmetric binary relation $P$  on a topological space $X$ and  $R=\left(P^c\right)^{-1}$ denote its reflexive hull.  Then, (also with relational notation in braces),  
\ben[{\nf (i)},  topsep=3pt]
\setlength{\itemsep}{-1pt} 
 \ml $R$ is  called {\it pseudo-transitive} if $(x,x')\in P, ~(x',y')\in R \mbox{ and } (y',y)\in P$  imply $(x,y)\in P$ for all $x,y,x',y'\in X,$         
 ($x\prec x'\preceq y'\prec y$  implies $x\prec y).$
 \lb{it: pseudotransitive}
 \ml  $P$ is called {\it separable} if there exists a countable subset $A$ of $X$ such that $(x,y)\in P$  implies  there exists $x'\in A \mbox{ such that }  (x,x')\in P \mbox{ and }  (x',y)\in P$, ($x\prec y \mbox{ implies } \exists x'\in A \mbox{ such that }  x\prec x'\prec y).$   \lb{it: separable}
  \ml $P$ is called {\it strongly separable} if there exists a countable subset $A$ of $X$ such that $(x,y)\in P$  implies  there exist $x',y'\in A \mbox{ such that } 	(x,x')\in P, ~ (x',y')\in R \mbox{ and }  (y',y)\in P,$ ($x\prec y \mbox{ implies } \exists x',y'\in A \mbox{ such that } 	x\prec x'\preceq y'\prec y).$    \lb{it: stronglyseparable}
 \ml $P$ has a  continuous dual-representation if there exist two continuous real valued functions $u$ and $v$ on $X$ such that $(x,y)\in P \mbox{ if and only if } u(x)<v(y)$ for all $x,y\in X,$ ($x\prec y \mbox{ if and only if } u(x)<v(y).) $  \lb{it: representation}
\een
\lb{df: chat}
\dff

\medskip

We can now present our final equivalence theorem. 
\vspace{3pt}

\thm
The following statements are equivalent for a binary relation defined on any  quasi-ordered topological space which contains more than two elements.    
\ben[{\nf (a)},  topsep=3pt]
\setlength{\itemsep}{1pt} 
\ml The space  is connected. \lb{it: c2}


\ml Any  strongly non-trivial and transitive relation with closed upper sections, and whose asymmetric part is negatively transitive with open upper sections, is complete and continuous. \lb{it: sccont}
\vspace{-15pt}


\ml Any two anti-symmetric, non-trivial and continuous relations on $X$ are either identical or inverse to each other. \lb{it: ei2}

\ml Any incomplete, non-trivial and transitive relation with closed sections, is fragile.
\lb{it: ge}

\ml Any incomplete, non-trivial and transitive relation whose asymmetric part has open sections, is flimsy.
\lb{it: sensitive}

\ml Any asymmetric relation with a continuous dual-representation is strongly separable.  \lb{it: ch1}

\ml Any asymmetric relation has a continuous dual-representation if and only if it is strongly separable, its dual is pseudo-transitive and its covering relations have closed sections. \lb{it: ch2}
\een
\lb{thm: cb}
\thmm
\vspace{-3pt}

Note that unlike Theorems 1 to 3,  continuity of the given relation  is not a standing hypothesis for Theorem 4, and thus the theorem can be read as an attempt to deconstruct the continuity postulate.  When we substitute the strong form of transitivity for continuity, we almost necessitate continuity. Again, the quasi-order assumption is only used in the proof of the implication \ref{it: ei2} $\Rightarrow$ \ref{it: c2}.

 The statement \ref{it: c2}$\Rightarrow$\ref{it: ei2} in the above theorem generalizes \citet[Theorem II]{ei41} by dropping completeness and transitivity; and  the statements \ref{it: c2}$\Rightarrow$\ref{it: ge},  \ref{it: c2}$\Rightarrow$\ref{it: ch1} and   \ref{it: c2}$\Rightarrow$ \ref{it: ch2} are due to \citet[Corollary 3]{ge13} and  \citet[Fundamental Lemma and Theorem]{ch87}, respectively. The statement \ref{it: c2}$\Rightarrow$\ref{it: sccont} shows that the continuity assumption in Theorem \ref{thm: c} can be weakened by strengthening the non-triviality assumption. Note that even though we did not explicitly assume negative transitivity in Theorem \ref{thm: c}, it follows from Theorem \ref{thm: ttransitivity}\ref{it: tippi} that the assumptions of the hypothesis of the theorem imply that $P$ is negatively transitive.

We conclude this section with the elementary observation that Theorems \ref{thm: c} and  \ref{thm: cb} can be collapsed into a single portmanteau result characterizing topological connectedness.  Such a portmanteau theorem, on its own, or in conjunction with Theorems \ref{thm: kc} and \ref{thm: 2cb},  can perhaps be read as an up-to-date survey   of the two-way relationship brought to light by topological assumptions on the binary relation and the set over which it is defined.\fn{We return to this issue in  Section 5 below; see Footnote~\ref{fn:rg3}, and the text it footnotes.   \label{fn:rg2}}


 \subsection{Sen's Deconstruction of the Transitivity Postulate}
 The results presented in this section can be introduced in two alternative ways:  (i) as an answer to questions that are naturally raised by the four equivalence theorems presented above, or (ii) as a generalization of  \cites{se69}  non-topological rendering of the results of Sonnenschein to the topological register. In terms of the first, let us consider what Theorems 1 to 4 bring to the table in terms of giving an underlying basis for the transitivity postulate. As already noted, Theorem 1 is primarily concerned with completeness,  and  Theorem 4  with separability, fragility, and flimsiness.  Theorem 2 does offer transitivity  as a conclusion but bundles it with completeness, while Theorem 3 uses completeness of the relation as a hypothesis. Natural questions then arise as to whether the dual conclusion can be unbundled, whether the completeness hypothesis can be dispensed with, and whether the role of connectedness pinned down. In short, this is to ask for some sort of minimal setting under which transitivity obtains, and  consistency is minimally unencumbered by decisiveness?   Theorems 5 and 6  below respond to these questions.\fn{Since we exclusively work with a topological structure, we do not investigate the implications of the added assumptions related to linear structure on the ES program. \citet{uz60} pioneered this line of research by showing that convexity of a complete preference relation defined on a convex subset of a topological vector space with closed upper sections and a transitive asymmetric part  implies that the relation itself is transitive. This result is reproduced and generalized by \citet[Theorems 5 and 6]{so65} and picked up by \citet{gku18}. \lb{fn: convexity}}

 But an alternative presentation of our results can be furnished in terms of what we are referring to as Sen's deconstruction of the transitivity postulate.  Towards this end, we refer the reader back to Sen's notation for the six different transitivity concepts $T, N\! P, I\! I, P\! P, I\! P, P\! I, $ already presented in Section 2. Referring to two transitivity conditions as {\it independent} if there exists a relation which satisfies one and  not the other, and  {\it interdependent} otherwise,  \cite{se69} has offered a non-topological examination  of the transitivity postulate.  In his Theorem I, he 
 provides  a synthetic treatment of the interdependence between the various transitivity conditions when the underlying primitive binary relation is complete;\fn{Note that to say that the symmetric part of a relation is complete renders the relation trivial, whereas to say that the asymmetric part is complete furnishes the contradiction that an element of the space is preferred to itself.    In \cite{se69}, $P\! P$ is the crucial transitivity condition and he found it convenient to give it a more usable name: ``quasi-transitivity''.  In this connection, also see the  subject index in \cite{se17} for additional discussion of his named concept.}   see his Figure 1, also  reproduced in \citet[p. 66]{se17}, and  as Figure 3(b) below.
Under completeness of $R,$ which Sen assumed,  $P\! P$ and any one of $P\! I,$ $I\! P$ and $I\! I$ imply $T,$ as the reader can see from Theorem 3b.\fn{For example, to see $P\! P$ and   $I\! P$  imply $T$ from Figure 3(b), simply note that     $P\! P$ and    $I\! I$  imply  $P\! I,$ which combined with $P\! P$ then implies $T.$  And so on for the other implications.}  The point of departure for our results is how Figure 3(b) unravels without the completeness postulate on $R$ (see Figure 3(a)), and how  most of these interdependence relations   can be recovered, and new relationships emerge,  if the underlying space and the preference relation is assumed to  satisfy some  suitable topological properties. As such, this inquiry falls very much  within the ES program, and in this   subsection we provide its  elaboration in this direction.\fn{There is a rich philosophical  literature on the discussion of non-transitivity, already referred to in Footnote~\ref{fn:ph1}. One may add that \citet{tu64} defends transitivity by using simple logical arguments and criticizes the experiments which argue evidence against transitivity. \citet{an93} criticizes Tullock for not considering ternary relation; also see also \citet{lu56}, \citet{fi70} and \citet{eo06}.  \label{fn:ph1}}

\smallskip

We begin with the following. 

\prp For any  binary relation $R$ on a set containing  at least four elements, and with  
   $I$ and $P$ denoting  its symmetric and asymmetric parts, the following statements are valid. 
\ben[{\nf (a)},  topsep=3pt]
\setlength{\itemsep}{-1pt}  \ml $T \Leftrightarrow P\! P \land P\! I  \land I\! P  \land I\! I.$ \lb{it: tr}
 \ml $N\! P\Rightarrow P\! P  \land P\! I  \land I\! P$. \lb{it: tnp}
    \ml $T$ is independent of $N\! P$.  \lb{it: trnp}
 \ml any subcollection of $P\! P, P\! I, I\! P, I\! I$ is independent of the remaining collection, severally and collectively. \lb{it: tindep}
\een  
\lb{thm: transitivity}
\prpp

\nt Proposition \ref{thm: transitivity}  implies that, in the absence of completeness, four of the transitivity conditions,  $P\! P, I\! P, P\! I$ and $I\! I,$ are independent of each other, severally and collectively; and  $T$ is independent of $N\! P.$ Therefore, the relationship between different transitivity concepts that Sen illustrates falls apart without the completeness assumption.

 We now turn to showing  that under suitable topological conditions, $P\! I$ together with $I\! P$ play an essential role for the completeness and the transitivity of $R,$ and  are thereby led to the notion of ``semi-transitivity''.\fn{We warn the reader of the difference in terminology; see Footnore~\ref{fn:semi} above.   \label{fn:semi2}}  But before the formalities, we present  a picture of the interdependence between the transitivity conditions without referring to the completeness assumption.  Panel\,(a) of Figure \ref{fig: sen} illustrates the interdependence  between different transitivity conditions for an arbitrary preference relation; panel\,(b)  (as already mentioned above) illustrates the relationship under the completeness assumption;  panel\,(c) remains with the incompleteness assumption but introduces continuity and connectedness assumptions; panel\,(d) includes completeness as well as continuity and connectedness assumptions. Label {CC} denotes that the underlying space is connected and that the binary relation defined on it is continuous. Label {CIC} keeps the continuity assumption of CC and replaces the connectedness of the space with  the connectedness of the sections of the symmetric part of the binary relation. To be sure, transitivity and completeness are totally different conditions and  completeness has implications on transitivity, but what our result brings out is that  transitivity has implication for  completeness.  As an example, negative transitivity of $\prec$ is equivalent to the following: if $x\prec y$, then $z\prec y$ or $x\prec z$ for all $z$ in the space. This conditions puts a limit on the level of incompleteness of the relation.\fn{As we shall have occasion to see below, all this supports the quotation from Wakker in the next section about the importance of judging the conditions in combination.}   
  
  \begin{figure}[ht]
\begin{center}
  \includegraphics[width=6in, height=4in]{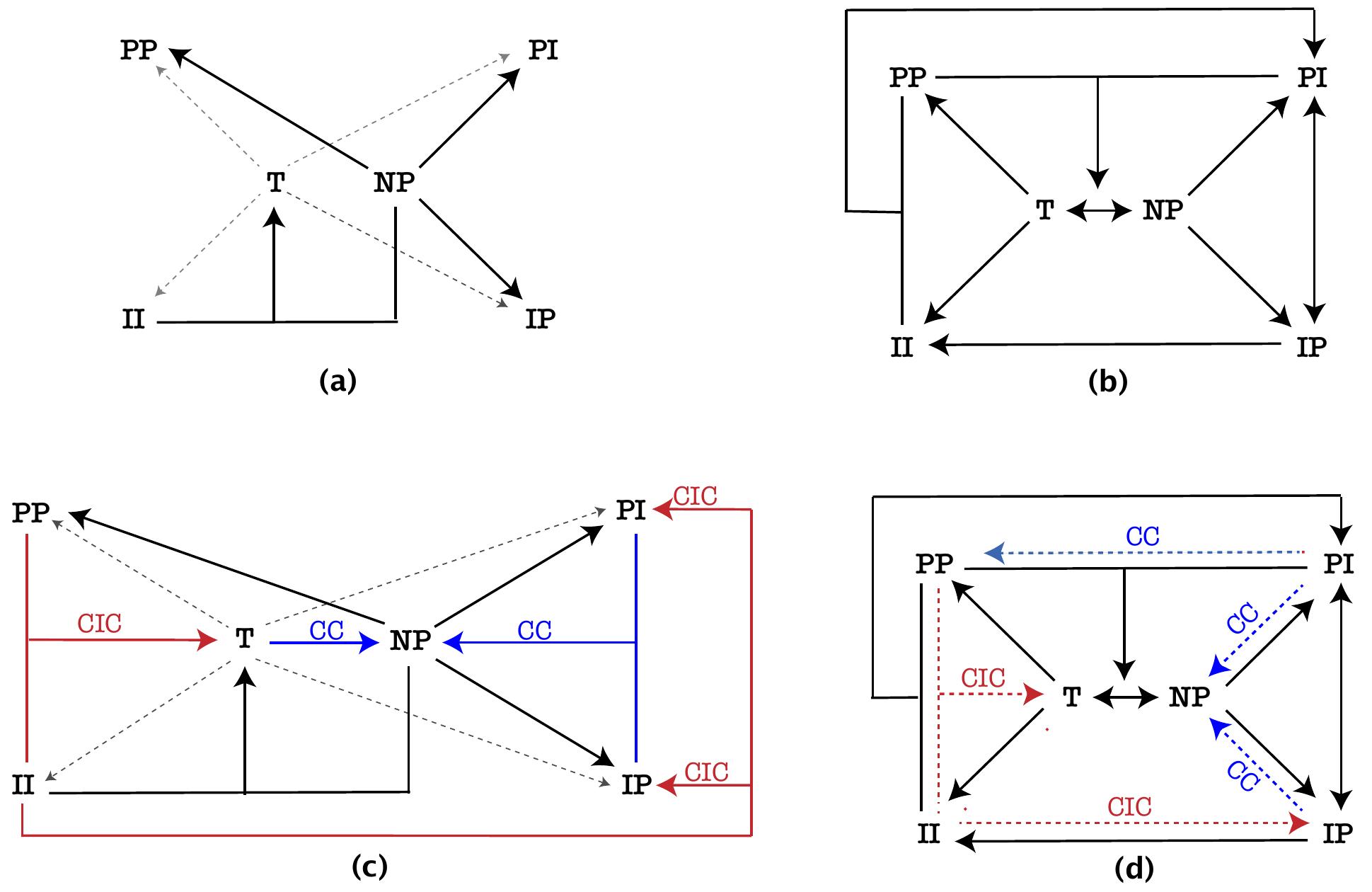}
    \end{center}   
    \vspace{-10pt}
    
\caption{The Interdependence of Different Transitivity Conditions}
\vspace{-5pt}   
   
  \lb{fig: sen}
\end{figure}

Theorems \ref{thm: ttransitivity} and \ref{thm: pbtransitivity} presented below are an attempt to  identify the most parsimonious setting to obtain transitivity, one that  has no reference to completeness at all.  As alluded to in the beginning of this subsection, they are 
 the next step from the point where Theorems \ref{thm: kc} to \ref{thm: cb} have brought us. They  extract what all can be said about transitivity without any other considerations: suitable topological assumptions on the choice set and on the preferences defined on it, allow us to  both reconstruct some of these relationships and also bring new ones to light. But in their single-minded concern with the postulate of transitivity, as in \cite{ei41},  \cite{so65, so67} and \cite{se69}, one of our findings, through decisive counterexamples, is that  equivalence has to be necessarily jettisoned.  As Proposition \ref{thm: cw} above, they limit themselves only to the forward direction.

%

\thm For any  continuous binary relation $R$ on a topological space, with  
   $I$ and $P$ denoting  its symmetric and asymmetric parts, the following statements are valid. 
\ben[{\nf (a)},  leftmargin=0.7\parindent, topsep=3pt]
\setlength{\itemsep}{-1pt} 
 \ml If the  space  is connected,  then  \lb{it: tippi}
  \ben[{\nf (i)}, leftmargin=0.4\parindent, topsep=-3pt]
    \setlength{\itemsep}{-1pt} 
    \ml semi-transitivity of $R$ is equivalent to negative transitivity of $P$  $(P\! I \land I\! P \Leftrightarrow N\! P)$, \lb{it: tippi1}
    \ml semi-transitivity of $R$ implies transitivity of $P$  $(P\! I \land I\! P \Rightarrow P\! P)$, \lb{it: tippi2}
    \ml transitivity of $R$ implies negative transitivity of $P$ $(T \Rightarrow N\! P)$, \lb{it: tippi3}
     \ml transitivity of $R$ is equivalent to its semi-transitivity and transitivity of $I$ $(T \Leftrightarrow P\! I \land I\! P \land I\! I)$. \lb{it: tippi4}
  \een
\vspace{-7pt} 
 
 \ml  If the sections of $I$ are  connected, then  \lb{it: tii}
   \ben[{\nf (i)}, leftmargin=0.4\parindent, topsep=-3pt]
    \setlength{\itemsep}{-1pt} 
    \ml transitivity of $I$ implies semi-transitivity of $R$  $(I\! I \Rightarrow P\! I \land I\! P)$, \lb{it: tii1}
    \ml transitivity of $R$ is equivalent to transitivity of $P$ and of $I$ $(T \Leftrightarrow P\! P\land I\! I)$. \lb{it: tii2}
\een  
\een
\lb{thm: ttransitivity}
\thmm

\nt Theorem~\ref{thm: ttransitivity}\ref{it: tippi} illustrates the essentially of semi-transitivity by showing that a continuous  and semi-transitive relation defined  on a connected space implies $P\! P$ and $N\! P$.  Whereas Theorem~\ref{thm: ttransitivity}\ref{it: tii} illustrates the essentially of the transitivity of $I$ by showing that for any continuous relation whose $I$ has connected sections, $I\! I$ implies semi-transitivity.  \citet[Theorem 4]{so65} uses connectedness of the sections of the indifference relation in order to obtain semi-transitivity of a complete and continuous relation whose symmetric part is transitive. All in all,  Theorem 5 provides a complete picture of the relationship between different transitivity conditions without using the completeness of the relation. This is all to say that under suitable topological assumptions, the transitivity of $I$ implies the remaining transitivity conditions. 


A natural question arises regrading the backward direction, and hence the possibility of equivalence. Such an equivalence can be written  in two parts, one for the setting where $X$ is connected, and the other when for the sections of $I$ are connected. The following simple example suggests that both are false. 

\exm {\nf Assume $X=\{a,b\}$ is endowed with the discrete topology. Then any relation is continuous and satisfy all of the six transitivity properties we use except $T$ and $I\! I$.  Hence conditions (i) to (iii) in part (a) hold. It follows from $P\! P, I\! P$ and $P\! I$ hold for any relation on $X$ that $T$ is equivalent to $I\! I$, hence condition (iv) holds.  However, it is clear that $X$ is disconnected.  
  The statement of the converse related to the sections of $I$ needs some elaboration. We can state the equivalence theorem for this part as follows: {\it ``For any  continuous binary relation $R$ on a topological space, the sections of its $I$ are connected if and only if $R$ satisfies {\nf (i)} and {\nf (ii)}."} The relation $I=X^2$ satisfies both conditions, however its sections are disconnected.   \qed
\lb{exm: sen}
}\exmm

We now move on to the formalities  required for the formulation of our interdependence result, and  follow \cite{so65} in  invoking two additional concepts: 
path-connectedness and the so-called Phragmen-Brouwer property, henceforth PBP. To be sure,  Theorem 5 above also connects  to  both Theorems 3 and 4 (without the completeness assumption of course) of Sonnenschein. 
In his Theorem 7, \citet{so65} uses path-connectedness of the upper sections of the relation in order to obtain semi-transitivity of a complete and continuous relation with a transitive symmetric part defined on a topological space with the PBP. Theorem 6, like Theorem 5,  provides a complete picture of the relationship between different transitivity conditions without using the completeness of the relation.  Note that unlike Sonnenschein,   we assume both upper and lower sections of the relation are path-connected. When a relation is complete, $I\! P$ and $P\! I$ are equivalent. It is easy to see that path-connectedness of the upper sections implies the latter property, but we need it for the lower sections to establish the former.  
  Noting that an  \lq\lq example of a class of sets for
which the PBP holds is the collection of all convex sets in Euclidean $n$-space," Sonnenschein nevertheless proves a special case of his result that \lq\lq  does not
rely on the rather deep PBP."\fn{See \citet[Footnote 4 and Theorem 7A]{so65} where the alterantive proof of the special case is justified on the grounds of furnishing \lq\lq an interesting technique."} A comprehensive discussion of the PBP, a property nothing if not elusive,   is furnished in \citet[Chapter II.4]{wi49}, and we refer the reader to it.\fn{ See, in particular, \cite[Theorems 4:12 and 9.3]{wi49}; also \cite{di84}, \cite{br06}, \cite{bc15} and their references for modern work on the property.   Note that the PBP  is not a strengthening of the connectedness assumption since any space with the discrete topology satisfies it, but rather a different separation property.}  

   For the purposes of this work we shall make do with the  following definitions.

\df
We call two non-empty subsets $A, B$ of a topological space $X$ {\it separated} if $\bar A\cap B=\emptyset=A\cap \bar B.$ We say that a set $A\subset X$ {\it separates} $x$ from $y$ if $x$ and $y$ lie in different components of $X\backslash A.$ We say that  $X$ has {\it Phragmen-Brouwer property} if for all separated open sets $A, B\subset X$ and all $x\in  A, y\in B,$ there exists  a connected subset of $X\backslash (A\cup B)$  which separates $x$ and $y.$ 
      A {\it path} in $A\subset X$ is a continuous function $s: [0,1]\ra A$, where $[0,1]$ is endowed with the usual topology.  
    A set $B\subset X$ is {\it path-connected} if for all $x,y\in B$ there exists a path  $s: [0,1]\ra B$ such that $s(0)=x, s(1)=y$. 
\lb{df: pb}
\dff

\nt   The following example illustrates a subset of Euclidean space which does not have the PBP.  It is well-known that in addition to a convex subset, the Euclidean $n$-sphere, $n>1$  satisfies  this property, and so by necessity, our example is based on $n=1.$

\exm {\nf Let $X=\{x\in \Re^2| x_1^2+x_2^2=1\}$ be endowed with the usual Euclidean topology. Let $x, y$ be two distinct points in $X$. Let $A$ and $B$ be the two components of $X\backslash \{x,y\}$. Then $A$ and $B$ are separated open sets. However, any connected subset of the subspace $\{x,y\}$ does not separate any $a\in A$ and $b\in B$.  \qed
\lb{exm: pb}
}\exmm

\nt With all the preliminaries behind us, we can finally present the  result showing  that the connectedness of the sections of $I$ in Theorem \ref{thm: ttransitivity}(b) can be replaced by the assumptions that $R$ has path connected sections and that the space has PB property.

\thm Let $X$ be a topological space with the  Phragmen-Brouwer property. Then for any  continuous binary relation $R$ on it, with $I$ and $P$ denoting  its symmetric and asymmetric parts, the following statements are valid.\fn{In a connected and locally connected space, path-connectedness of the sections of a complete relation $R$  implies that $I$ has connected sections. Moreover, connectedness of the sections of $I$  in a connected space implies that $R$ has connected sections, which is weaker than path-connectedness. See \citet[Theorem 4.5 on p. 49 and Theorem 9.9 on p. 20]{wi49} for the proofs.}  
   \ben[{\nf (i)},  topsep=-4pt]
    \setlength{\itemsep}{-1pt} 
    \ml If $R$ has path-connected upper sections, then transitivity of $I$ implies $P\! I$  $(I\! I \Rightarrow P\! I)$. \lb{it: tpb1}
    \ml If $R$ has path-connected lower sections, then transitivity of $I$ implies $I\! P$  $(I\! I \Rightarrow  I\! P)$. \lb{it: tpb2}    
    \ml  If $R$ has path-connected sections, then transitivity of $R$ is equivalent to transitivity of $P$ and of $I$ $(T \Leftrightarrow P\! P\land I\! I)$. \lb{it: tpb3}
  \een 
\lb{thm: pbtransitivity}
\thmm

Again, a natural question arises regarding the backward direction, and hence the possibility of equivalence. 
We can formulate an  equivalence conjecture  as follows: {\it ``A topological space has the PBP if and only if any continuous binary relation $R$ on it satisfies {\nf (i)} to {\nf (iii)}."}  Unlike connectedness, working with a topology with a lot of open sets will not yield a counterexample to PBP; in fact we have to work with a topology with few open sets. But we cannot also work with a very poor topology in order to obtain a counterexample to the property since there may not be enough separated sets.  In any case, the following example shows that the converse of Theorem \ref{thm: pbtransitivity} is false. 
\smallskip

\exm {\nf  Let $X=\{a,b,c,d\}$ be the choice set which is endowed with the topology  $\tau=\{\emptyset, \{a\}, \{c\}, \{a,c\}, \{a,b,c\}, \{a,c,d\}, X\}$. Note that $\{a\}$ and $\{c\}$ are separated open sets. Moreover,  $\{b,d\}$ is disconnected and $\{a,b,c\}, \{a,c,d\}, X$ are connected.  Then any connected subset of  $\{b,d\}$ does not separate $a$ and $b$. Therefore, $X$ does not have the Phragmen-Brouwer property.  

Next, we show that any continuous binary relation $R$ on $X$ satisfies conditions (i) to (iii) of Theorem \ref{thm: pbtransitivity}. Note that if we prove conditions (i) and (ii), then Proposition \ref{thm: transitivity} implies condition (iii). To this end, assume that the symmetric part of the  relation is transitive. We first show that path-connectedness of the upper sections of $R$ implies that $P\! I$ holds.  
 Note that the collection of closed sets in $X$ is $\{\emptyset, \{b\}, \{d\}, \{b,d\}, \{b,c,d\}, \{a,b,d\}, X\}$. It is easy to show that among the closed sets, only $\{b,d\}$ is not path-connected. Therefore, the upper sections of $R$ must belong to the collection $\{\emptyset, \{b\}, \{d\}, \{b,c,d\}, \{a,b,d\}, X\}$. Since  $I(x)=R(x)\cap R^{-1}(x)$ for all $x\in X$ and $R$ has closed sections, therefore $I(x)$ belongs to the collection  $\{\emptyset, \{b\}, \{d\}, \{b,d\}, \{b,c,d\}, \{a,b,d\}, X\}$ for all $x\in X$. Note that all members of this collection is connected except $\{b,d\}$. Then for any relation $R$ on $X$ with $I(x)\neq \{b,d\}$  for all $x\in X$, it follows from  Theorem \ref{thm: ttransitivity}(b) that $R$ is semitransitive, hence condition (i) holds.  

We now show that if a continuous binary relation $R$ on $X$ has path-connected upper sections and a transitive symmetric part, then $I(x)\neq \{b,d\}$ for all $x\in X$. Assume that  for some $R$ there exists $x\in X$ such that $I(x)=\{b,d\}$. Then the transitivity of $I$  implies that $x=b$ or $x=d$. Let $x=b$. Since  $R(b)$ is closed, path-connected and  $\{b,d\}\subset R(b)$, therefore   $R(b)$ is equal to one of $\{b,c,d\}, \{a,b,d\}$ and $X$.   
 Let $R(b)=\{b,c,d\}$. Then it follows from $R$ has closed sections and  $I(b)=\{b,d\}$ that  $R^{-1}(b)$ is equal to either $\{b,d\}$ or $\{a,b,d\}$. In both cases, $P(b)=\{c\}$. Since $b\in P^{-1}(c)$ and $P$ has open sections, therefore $P^{-1}(c)$ must contain $c$, which furnishes us a contradiction. Now let $R(b)=\{a,b,d\}$. Then $R^{-1}(b)$ is equal to either $\{b,d\}$ or $\{b,c,d\}$. In both cases,  $P(b)=\{a\}$. Since $b\in P^{-1}(a)$ and $P$ has open sections, therefore $P^{-1}(a)$ must contain $a$, which furnishes us a contradiction. Finally, let $R(b)=X$. Then $R^{-1}(b)=\{b,d\}$. Hence,  $P(b)=\{a,c\}$. Since $b\in P^{-1}(a)$ and $P$ has open sections, therefore $P^{-1}(a)$ must contain $a$, which furnishes us a contradiction. Therefore,  $x\neq b$. Analogously, the case $x=d$ yields contradiction.

An analogous argument shows that  path-connectedness of the lower sections of $R$ implies that $I\! P$ holds.    Therefore, the example is complete.     \qed

\lb{exm: pbc}
}\exmm


Theorems \ref{thm: ttransitivity} and \ref{thm: pbtransitivity} are in-line with the four theorems we present in the previous section. In the forward direction of those theorems we show that under suitable topological assumptions on the choice space and the preferences defined on it, non-triviality and weak forms of transitivity imply full transitivity as well as the completeness of the preferences.  The proposition above and these two theorems present the relationship among different transitivity conditions without referring to the completeness assumption. In particular, assertions \ref{it: tippi} and \ref{it: tii}  in Theorem \ref{thm: ttransitivity} are analogous to assertions \ref{it: ess} and  \ref{it: es} of Theorems 1 and 2 above.

 We conclude this section by observing that Theorems  \ref{thm: ttransitivity}  and \ref{thm: pbtransitivity} have implications for the topological structure of the graph of a binary relation. A binary relation $P$ on $X$ is said to have {\it open graph} if $P$ is open in the product topology on $X\times X.$ \citet[Theorem 2 and Corollary]{bpr76} proved that an asymmetric and negatively transitive binary relation on a topological space has open sections if and only if it has open graph.  The two theorems above  show  that we can weaken negative transitivity by strengthening the topological assumptions on preferences and on the space.\fn{\citet{ge15} shows that reflexive, transitive and additive binary relation with closed upper sections has a closed graph, and hence,  convexity of a relation has implications on its continuity; see Footnote \ref{fn: convexity} for further references and discussion on convexity.}


\section{Applications of the Theory: A Brief Excursus} \lb{sec: app}
Results in  pure theory are to be judged not only for their application to real-world problems, but also by  new light they cast on the earlier theoretical results themselves:   they  allow, indeed enable and empower, one to really see what was already seen before. In this section, we do this through the help of two lenses:  {\it redundancy} and  {\it hiddenness} of hypotheses in a rather   extensive antecedent literature. The postulates of completeness and transitivity serve  as hypotheses to conclusions concerning results (i) on the representation of a preference relation by a real-valued (utility) function, (ii) in neoclassical consumer theory,  (iii)  in Walrasian and Cournot-Nash equilibrium theory, and (iv)  in social choice theory; and it stands to reason that our six results and the examples would have some impact on these results.   {\it Redundancy,} the removal of hypotheses unnecessary for a conclusion,  is simply bringing Occam's razor into play, a procedure with a long and rich lineage in mathematical investigation. It goes into what one means by generalizing a theorem. {\it Hiddenness}  is somewhat more subtle, and more far-reaching. Rather then saying that a hypothesis can be eliminated from an assertion, it says that it is already incorporated in other hypotheses already assumed.    We say that a particular  assumption is hidden in a theorem when it is already implied by the others in the battery of assumptions constituting the hypothesis of the theorem.   As such, showing that it can be removed does not lead to a generalization of the result, but merely to obtaining its clearer formulation and a more parsimonious  expression. But it also includes making an implicit assumption or drawing out its fuller meaning of an assumption in the sense of uncovering the implications that are hidden in it.\fn{A celebrated example of the former is \cite{ma52}; also see Footnote~\ref{fn:ma2} below. An example of the latter is a reconsideration of Shafer's non-transitive consumer that we present in Section 4.2, and of Schmeidler's existence theorem in Section 4.3; also see Footnote~\ref{fn:sc} below. What we do not mean by the word {\it hiddenness}  is the sense that is given to it by \citet{ty00}. Their usage is orthogonal to ours, and perhaps also to {\it redundancy}: they refer to an assumption being \lq\lq hidden" as one that is explicit but essential in the sense that eliminating it would require additional hypotheses to obtain the same consequences. \label{fn:ma1}}   In this section, we work this distinction though the antecedent literature listed above.

 With this background we return to \cites{wa88a} assertion that technical assumptions by themselves  \lq\lq do not entail obscurity and are not very bothersome."  His text continues beyond our truncated citation. 
 
 \bqu  However, when one requires a list of conditions, then one should not judge each condition separately, but one should judge the conditions in combination. We give an example  where, paradoxically, each individual condition involved is not falsified by the observations, but the combination of the conditions is falsified. As it turns out in this example, continuity, {\it in the presence of other conditions,} may have empirical meaning.\fn{The italics are the author's, and the example is  \citet[Example 7.3]{wa88a}. For further discussion on the empirical implication of technical assumptions, see \citet[pp. 107-108]{pf71} and \citet[pp. 83-84]{na85}.} It is very bothersome that usually the exact empirical meaning of simplifying non-necessary conditions such as continuity is unclear.
\equ

 \nt However, rather than Wakker's example, focus simply on a decision-maker with a  choice set of $\Re_+,$ and who, when faced with the options $\{0, 1\},\;  \{1, 2\}, $ and $ \{2, 0\},$ chooses $1, 2$ and $0$ respectively. Since his choices do not falsify  anti-symmetry and non-triviality, and, irrespective of the number of additional but finite observations, cannot falsify continuity and exhibit non-transitivity, an appeal to the implication \ref{it: c}$\Rightarrow$\ref{it: eg} in Theorem \ref{thm: c}, and a presumption of anti-symmetry would falsify continuity.  And this is all that is being asserted about continuity, in the presence of other conditions, having  both {\it behavioral} and {\it empirical} implications. 
 
Finally,  before taking our theorems to the antecedent literature, we take notice   of three rather current references  that engage with Schmeidler's theorem in directions   somewhat oblique  to ours.   \citet[Proposition 10]{go18} presents a result that Schmeidler's theorem as a special case, but rather than the necessity and sufficiency of  the property, resets it on maximal $\succeq$-domains, maximal on which the relation is complete. In a somewhat similar vein,  \citet{co18} work on the structure of an incomplete and non-transitive preference relation by identifying and examining the properties of largest transitive sub-relation of it.  They assume partial continuity of the relation, and are well aware that full continuity would  otherwise imply completeness. However, they make no connection to the work of Eilenberg, Sonnenschein and Sen.   \citet{no18} work with a preference structure consisting of two relations, one transitive but incomplete and the other complete but non-transitive. In this interesting paper,  they redo notions of maximality, social choice and  decision theory. It would be interesting to see whether the results reported above have any relevance to these lines of work. 

We now turn more sure-footedly to papers that are directly impacted by our theorems.  However, it is important to keep in mind that it is not our intention  to give a comprehensive catalogue of each and every result to which our six theorems can be fruitfully applied. We content ourselves with laying out excursionary   directions, and  leaving it to the reader to pursue in more detail the direction that interests him or her.  Furthermore, we  emphasize the ``forward'' direction in our applications.

\subsection{Real-valued  Representation of Preferences}
In \citet[Section 5]{bb82}, the authors present a representation theorem that they ascribe to Debreu, Eilenberg and Rader.\fn{Also see \cite{wa88b}.  For comprehensive treatments, see   \citet{bm95}, \citet{me98}, \citet{hd18}  and their  references.}  We 
begin with  the simple version due to \citet[6.1]{ei41}'' in his seminal paper.\fn{Note that \citet[2.1]{ei41} observes the hiddenness of transitivity. Moreover, our insertion of non-triviality in the statement of Eilenberg's theorem do not do violence to his original statement because of the consideration emphasized in Footnote~\ref{fn:triviality} and in the text it footnotes. 
}
\bqu
{\it Every non-trivial, anti-symmetric, complete, transitive and continuous preference relation on a connected and separable topological space  has a continuous utility representation.} 
\equ

%
\nt It follows from implication \ref{it: c}$\Rightarrow$\ref{it: eg} in Theorem \ref{thm: c} that both completeness and transitivity follow from the remaining assumptions of his theorem, hence they are {\it hidden} assumptions of his statement.  Therefore, we can equivalently state his theorem by dropping both completeness and transitivity assumptions.  
 Moving on to the variant of Eilenberg's result in \citet[Theorem I]{de54}, one that dispenses with his anti-symmetry assumption,  it is already well-understood that  as the consequence of the theorems of   of Sonnenschein and Schmeidler,  either completeness or full transitivity is hidden in Debreu's assertion. The point that we wish to emphasize here is that as a consequence of  \ref{it: c}$\Rightarrow$\ref{it: ess} in Theorem \ref{thm: c},  not one but that {\it both} completeness and full transitivity are hidden in Debreu's theorem.  Moreover, \citet[p. 42]{wa89}, as well as \citet[pp. 65-66]{fi72},  observe the hiddenness of either one of completeness and transitivity for the existence of a utility representation, and our point is  again that {\it both} of these assumptions are hidden in this line of the  literature.\fn{For additional references, the interested reader may see, for example,  \citet{de60}, \citet{go68}, \citet{bm95} and \citet{vi03}.}
 
 This being said, it is important to emphasize that applications that  work with  strict preference relations as their primitive, such as  \citet{pe70} and \citet{ms76},\fn{While Peleg's representation provides only froward direction representation result for incomplete preferences, Majumdar-Sen's approach provides both directions.}   or those that simply drop both completeness and transitivity such as \citet{no16},  also do not fall under the ambit of this paper.   The same is true for applications relying on linear structures:  they do not fall under the exclusively topological rubric of this paper, and  require different mathematical techniques and tools, and we investigate such structures elsewhere; see \citet{gku18}.\fn{This literature stems from   
 the classic work of \citet{hm53}, where the topological structure is restricted to that on the unit interval, and the overall thrust is algebraic.}   But again, just because the setting is one of 
 uncertainty, it does not mean that the results are necessarily out of the bounds of our treatment elaborated in this paper.  In those that rely on  the  continuity assumptions which we use in this paper, as, for example \citet[Theorem 1.3]{ka14},  the  implication \ref{it: c}$\Rightarrow$ \ref{it: ess} in Theorem \ref{thm: c} yet again implies that both completeness and full transitivity are hidden assumptions. 

There has been a surge of recent work on decision theory without the completeness postulate in a setting where the choice set is convex, and therefore connected.   Our results,  specifically Theorem \ref{thm: c}, then   suggest that any result for a setting with incomplete preferences, must of necessity weaken 
 the continuity assumption. In this line of literature, most papers drop the assumption of open sections; see for example \citet{gmms03}, \citet{dmo04} and \citet{ev14}. The implication \ref{it: c2}$\Rightarrow$\ref{it: ge} in Theorem 4 then yields the consequence  that   the preference relation must be  fragile! To be specific,   \citet[Expected Multi-Utility Theorem]{dmo04} satisfies all of the assumptions of the  implication \ref{it: c2}$\Rightarrow$\ref{it: ge} in Theorem 4, and hence the preference relation the authors work with is fragile, i.e., there exist $(x,y)\in P$ such that every open neighborhood of $(x,y)$ contains non-comparable elements. We can, moreover, use the other assumptions of their theorem in order to obtain  more information about the structure of their incomplete preferences as follows. Their theorem implies that there exists a closed and convex set $\mathcal U$ of continuous utility functions that represents $R$.  Fragility implies that for all $u\in \mathcal U$,  $u(x)<u(y)$  and any open neighborhood of $(x,y)$ contains $(x',y')\in V$ such that $u'(x')> u'(y')$ (note that weak inequality contradicts fragility) for some $u'\in \mathcal U$. Since the space of utility functions is endowed with the sup-norm topology, it is easy to show that there exists an open neighborhood $U$ of $u'$ such that $v'(x')> v'(y')$ for all $v'\in U$. Hence, there is an open set of utility functions that rank $x'$ and $y'$ opposite of the ranking of $x$ and $y$.

\subsection{Shafer's Non-Transitive Consumer} 
In a direction initiated by  \citet{so71},    
  \citet{sh74} re-works the neoclassical theory of demand for a    consumer with incomplete and possibly non-transitive preferences. Our results impact his work not by bringing out any redundancies, but  bringing in  what non-transitivity assumption fully entails in the light of his other hypotheses. We  show that if the preferences of Shafer's non-transitive consumer satisfies a little bit of consistency, then it is fully destructive of {\it all} non-transitivity. This is to say that ,  if any of the four dis-aggregations  $I\! I, I\! P, P\! I$ and $P\! P$ of  $R$ holds, then the remaining three also hold  under his assumptions.  And so Shafer's  non-transitive consumer has to be, by necessity, a  {\it fundamentally} non-transitive agent.  We elaborate this claim in the following paragraph; it seems to have been missed in the literature. 

Shafer assumes that the consumer has a complete, continuous and strictly convex preference relation $R$ on $\Re_+^n$. Strict convexity implies that $R$ is non-trivial and has path-connected upper sections, and  convexity of $\Re_+^n$ implies that the choice set is connected.  We can now develop the argument for our assertion under three cases. First, if $R$ satisfies $I\! I,$ then by  Theorem \ref{thm: pbtransitivity}\ref{it: tpb1},  $I\! I$ implies that $P\! I$ holds. Then it follows from completeness of $R$ and \citet[Theorem I]{se69}, which we illustrate in Figure \ref{fig: sen}, that $I\! P$ also holds. Hence $R$ is semi-transitive and its symmetric part is transitive. Then the implication \ref{it: c} $\Rightarrow$ \ref{it: ess} in Theorem \ref{thm: c} implies that $R$ is transitive. Second, assume $R$ satisfies $P\! I$. Then Sen's theorem implies that $I\! P$ and $I\! I$ hold. As above,  the transitivity of $R$ follows from Theorem \ref{thm: c}. The proof is analogous if $R$ satisfies $I\! P$. Finally, if $R$ satisfies $P\! P,$ the transitivity of $R$ follows from \citet[Theorem 5]{so65}.

\citet{ge10} re-works  Shafer's theory of the non-transitive consumer by also eliminating the completeness postulate and  weakening the continuity assumption.    Under this sparser structure, a consumer can satisfy some form of transitivity  without destroying all forms of transitivity.  A consumer can admit consistency in some registers without  consistency in all.  However, what emerges is the essentiality of the transitivity of $II$:  even if we drop completeness in Shafer's model, under full continuity assumption (the sections of $R$ are closed and of $P$ are open) and the convexity assumption of Shafer, our Theorems 2 and 6 that if $I\! I$ holds then the relation $R$ has to be transitive. Therefore, Shafer's non-transitive consumer necessarily violates the transitivity of $I$ even if she has incomplete preferences.\fn{This connects us to \cites{lu56} semi-order, which is further elaborated in \cite{fi70}, a connection that we hope to explore in future  work.} 

We conclude this section with the observation that the points that we make in the first two paragraphs above could already  have been made in 1965 drawing only on the results of \cite{so65}; as such,  they do not require the full power of our results. In any case, they shed new light on what {\it hiddenness} may entail.\fn{We also single out in this connection, \citet[Proposition 1]{mo96}. This  claims that transitivity  on every closed interval of a preference ration defined on a linear space follows from a weak convexity assumption, and as such, hidden by the convexity postulate.  Since we are limiting ourselves to the topological register, we hope to engage this claim elsewhere; also see footnote \ref{fn: convexity} for further references and discussion regarding the  convexity assumption.}

\subsection{Walrasian Economies and Normal-Form Games}

In this section, we first illustrate the hiddenness of completeness and full transitivity assumptions in the results on the existence of an equilibrium in Walrasian economies and in normal form games. Then we show how these observations carry over to the economies with indivisibilities.

The classical equilibrium existence results in Walrasian economies assume that the choice set of each consumer is a convex subset of the Euclidean space  and that each consumer has a complete, transitive and continuous preference relation on the choice set; see for example  \citet[Theorem I]{ad54} and \citet[Theorems 5 and 8]{de82}. Moreover, one of the following two properties is assumed:  {\it monotonicity}  or {\it non-satiation}. Since the former implies more is better for each consumer and the latter that no consumer has a best element in her consumption set, the  preference relation of each consumer is non-trivial. Since convexity of the choice set implies its connectedness,  all of the assumptions of the implication 
\ref{it: c}$\Rightarrow$\ref{it: ess} in Theorem \ref{thm: c} are satisfied. Hence both of the completeness and full transitivity postulates are hidden for the existence of a Walrasian equilibrium.  However  hiddenness is also present in many other results in this literature.     We invite the reader to check out theorems on the existence of a Walrasian equilibrium with continuum of agents, or externalities, or public goods, or infinite dimensional commodity spaces; see for example \citet{mz91}, \citet{ks02} and \citet{mc05}.  However, one can go beyond hiddenness to make  points akin to that made above regarding Shafer's non-transitive consumer.  If, for example,   the continuity and transitivity assumptions  made in \cite{sc71} are also made in \cite{sc69}, the existence of 
competitive equilibria in markets with a continuum of traders and incomplete preferences follows as a straightforward consequence of Aumann's existence theorem, and hardly requires an additional independent proof.

 In finite games, \citet{na50}  assumes each player has a finite number of pure strategies and that the preferences of each player defined on the set of all probability distributions on the set of pure strategies  satisfy the axioms of the classic expected utility representation theorem.\fn{As shown by \cite{ma52}, \citet{vm47} do not state the independence axiom in their axiomatization. \citet{na50b}  and \citet{ma50} independently provide complete axiomatization of expected utility; see \cite{bl16}, and also Footnote~\ref{fn:ma1} above. \label{fn:ma2}}   Even though he assumes a weaker continuity assumption than we use, in the presence of other assumptions it implies our stronger continutiy assumption; see \cite{du11}. Therefore, both completeness and full transitivity assumptions are hidden in Nash's results. 
For games with continuum of actions, \citet[Theorem]{de52} follows a different path (which eliminates randomization) and proposes a generalizes Nash's theorem by assuming the choice space is a convex and compact subset of a Euclidean space and the preferences defined on it are complete,  transitive, continuous and convex.\fn{We summarize the assumptions of the version of Debreu's theorem re-stated in \citet[Lemma]{ad54}.} Then, as illustrated above, the implication 
\ref{it: c}$\Rightarrow$\ref{it: ess} in Theorem \ref{thm: c} suggests the hiddenness of both completeness and full transitivity assumptions.\fn{Note that without non-triviality, the existence of an equilibrium is triviality, hence the non-triviality assumption is non-restrictive.}  Therefore, we can, equivalently re-state Debreu's theorem by dropping the completeness and weakening the transitivity assumption.

Although connectedness, or more precisely convexity, of the choice space is a common assumption in economics, an important class of  models which study markets with indivisibilities naturally assume disconnected choice spaces; see for example \citet{di71}, \citet{br72}, \citet{ma75, ma77}, \citet{ky81} and  
  \citet{th11}. In these models, there are $\ell$ indivisible goods and 1 divisible good. For illustration, assume $\ell=1$.  In particular, assume the consumption set of a consumer $X=\Re_+\times \Ze_+$ consists of money, which is perfectly divisible, and an indivisible good. Let $\succeq$ denote the preferences of the consumer on $X.$ Let $\sim$ and $\succ$ denote the symmetric and asymmetric parts of $\succeq,$ respectively. The following assumptions are standard in these models.

\begin{figure}[htb]
\begin{center}
 \includegraphics[width=4.5in, height=3in]{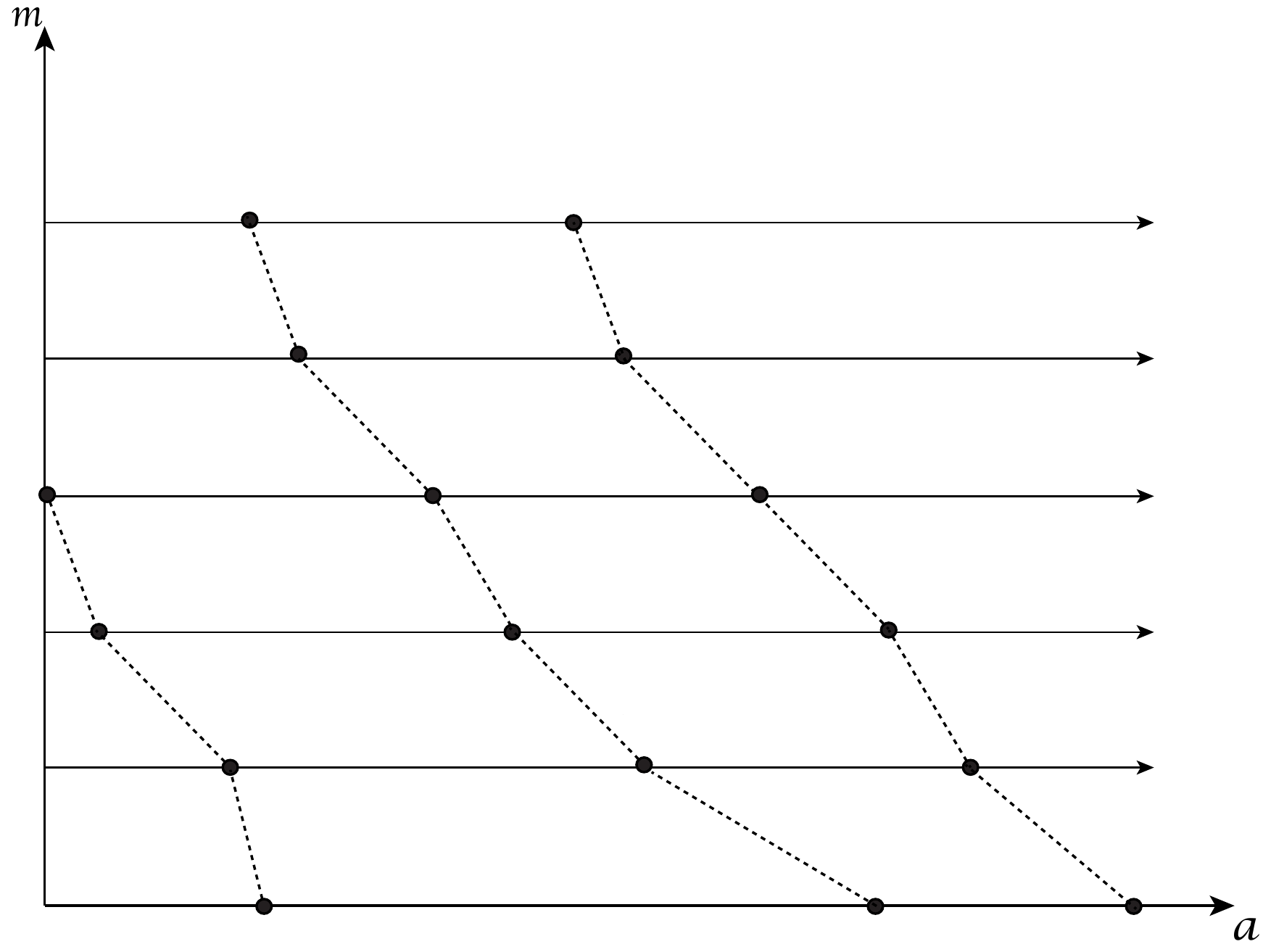}
\end{center}  
\vspace{-15pt}   
    
\caption{Economies with Indivisible Commodities}
\vspace{-5pt}   
   
  \lb{fig: indivisible}
\end{figure}
\smallskip

\noindent {\it 
\ben[leftmargin=1.15\parindent, topsep=3pt]
\setlength{\itemsep}{1pt} 
\ml[{\nf (A1)~}] For all $a\in \Ze_+,$ $(m, a)\succ (m', a)$ whenever $m>m'.$  \lb{it: id_m}
\ml[{\nf (A2)~}] For all $a, a'\in \Ze_+,$ there exist $m, m'\in \Re_+$ such that $(m,a)\sim (m',a').$ \lb{it: id_pc}
\ml[{\nf (A2$'$)}] For all $(m,a), (m',a')\in \Re_+\times \Ze_+,$ there exists $\lambda\in \Re_+$ such that $(m+\lambda ,a)\succeq (m',a').$
\een}
\medskip

\noindent Assumption (A1) is called {\it strict monotonicity in the divisible commodity}, (A2) {\it possibility of compensation} and (A2$'$) {\it overriding desirability of the divisible commodity}. The latter two assumptions are substitutes and used interchangeably in the literature.  Figure \ref{fig: indivisible} illustrates an economy with indivisible goods which satisfy these standard assumptions.

 \citet[Theorem 4.11]{br72} shows that if a complete, transitive and continuous preference relation on $X$ satisfies Assumptions (A1) and (A2) along with the assumptions that the consumption set and the preferences satisfy some additional topological and linear structure properties,\fn{See Assumptions 2.1 (excluding (e) and (f)), 2.6, 2.7 and 2.8 in \citet{br72}.}  then there exists an approximate equilibrium.  We now illustrate that completeness and full transitivity are hidden in the hypothesis of Broome's theorem. Consider a semi-transitive and continuous preference relation $\succeq$ on $X$ whose symmetric part is transitive.  It is clear that $X$ is disconnected -- for each $a\in \Ze_+,$ $C_a= \Re_+\times \{a\}$ is a component of $X.$ It follows from (A1) that there are strictly comparable elements within each component and from any of (A2) or (A2$'$) that there are weakly comparable elements across different components. Therefore, (A1) along with any of (A2) or (A2$'$) imply $\succeq$ is componentwise non-trivial. Therefore, Proposition \ref{thm: cw} implies that both completeness and full transitivity are hidden assumptions in Broome's theorem.   

In this line of research, \cite{ma74}, \cite{ss75} and their followers work with incomplete and non-transitive preferences, and assume only a strict preference relation, one with open graph or open sections. Hence, our results do not have implications for their work, as indeed, they also do not impact the literature on  games with discontinuous preferences pioneered by \citet{dm86} and \citet{re99}, or the reformulation of Cournot-Nash equilibria presented in  \cite{ks90}.


\subsection{Other Potential Applications}  Our final tripartite subsection is as much an invitation to the reader to apply the  results reported above, as it is a laying-out of directions for further work. We begin by asking whether the added specification of finite choice sets allows, if not a sharpening of the results, an opening into other productive directions. We then move on to other potential applications to classical social choice theory and to graphs and  networks.   

\subsubsection*{Finite Choice Sets}
 Eilenberg's remarkable paper notwithstanding,  modern decision and social-choice theory has focused on  compactness rather than on connectedness, and has limited itself to a finite setting for an exploration of ideas. It is thus natural to ask whether the results reported in Section 3 of the paper can be  extended to provide sharper results simply by seeing a finite set of $k$  alternatives as  a $k$-connected set. One rather obvious impediment to this is the fact that the hypotheses, say  of  Theorem 1,  require $P$ to have a strict relationship within every component, a property precluded by singletons in a  finite setting.  The question then reduces to whether indifferent alternatives  can simply be \lq\lq factored out"?    

We present a result that gives the negative answer to this question:  we show that one cannot define a non-trivial, complete, semi-transitive and continuous binary relation on a finite and connected topological space.

\prp
If $R$ is a complete, semi-transitive and continuous binary relation on a topological space $X$, then the quotient topology on $X|I$ is Hausdorff. 
\lb{thm: discrete}
\prpp

\noindent For a finite choice set, this proposition shows that the quotient topology with respect to the symmetric part of a continuous binary relation is discrete, hence for any connected set in the original space, all of its the elements must be indifferent to each other.  As such this direction is a dead end and substantial rethinking is needed, perhaps along the lines of the literature stemming from the application of convex geometry recently developed  in the imaginative contribution of \cite{rr15}; see \cite{ej85} for an early survey.   

\subsubsection*{Collective Choice and  Social Welfare }

Moving on to the substance of the theory itself, the  last two decades have seen a substantial maturing of the theory of  social choice and welfare; see \cite{ass95}, \cite{ass02}, \cite{fsy},  \cite{fb13}, \cite{su16}, \cite{se17} and their references.

Leaving aside the rich philosophical and technical subtlety of this  literature, the point is that it involves establishing the consistency, aggregation if one prefers to be more specific, of  two types  of binary relations: preferences of individuals and those of the group that those individuals constitute. The question reductively reduces to asking whether group and social preferences are  ``nice" when the individual  preferences are ``nice"? And to be sure, the  adjectives formalizing these valorizations  necessarily reduce to postulates such as completeness and transitivity, and thereby bring into play the six theorems that we present above. To get down to specifics, one can delineate how the the {\it hiddenness} and {\it redundancy} criteria   impact the theorems of   \citet{ha55} and \citet{sa81}, or factor into the recent exercise that \citet{ma10}  has carried out also in a purely topological register. 

Some what more obliquely, but perhaps even more promisingly,   \cite{br09}, propose \lq\lq replacing the standard revealed
preference relation with an unambiguous choice relation: roughly, $x$ is (strictly) unambiguously chosen over $y$ (written $xP^*y$) iff y is never chosen when x is available." They write

\bqu If one thinks of $P^*$ as a preference
relation, then our notion of a weak generalized Pareto optimum
coincides with existing notions of social efficiency when consumers have incomplete and/or intransitive preferences ...  . [t]hough $P^*$ need
not be transitive, it is always acyclic, and therefore suitable for
rigorous welfare analysis.\fn{See \citet[pp. 75-76]{br09}; also their Theorem 7 which they see as following directly from standard results of    Fon,  Mandler, Otani,  Rigotti and Shannon.} 
 \equ 

\noindent As such, the authors do not  take a stand on a particular story about why choices violate, for example,  $WARP$, the weak axiom of revealed preferences; but by subsuming  behavioral economics as theories that violate WARP, present explorations that surely fall within the rubric of the ES program.

\subsubsection*{Communication  Networks}
Reverting to finite choice sets and to convex geometry and its implications for partially ordered sets and in graphs, we close this subsection on potential applications by drawing the reader's attention to the  analysis of networks, be they social, economic,   political or anthropological.   This is  a very active field of microeconomic theory; see, for example, \cite{ja08} and his references.  

 A simple application of our result on $k$-connectedness to information transmission networks, as culled from \citet[ Chapter 7 and Section 13.2 ]{ja08} and  \cite{ne06}, draws on the observation that transmission of information in small communities,  villages and such, can be seen as a network whose node can be seen as a  {\it component} in the technical sense that we have given to the word in the work reported in this paper.  Completeness and transitivity of  communication relationships, in the sense of a node $x$ directly or indirectly communicating with the node  $y$,  seems formalizable and directly relevant.  The continuity of the communication relationship in terms of the distance between  villages,  the topology being  defined by this distance.  In this simple formulation, 
 let $X\subset \Re^2$ be the set of individuals which is endowed with the Euclidean topology. The distance between different points represent the distance of the individuals.
  Let 
   each node (village) denoting a component of $X$. Define a binary relation $\succeq$ on $X$ as follows: $x\succeq y$ if and only if $x$ (directly or indirectly) transmits information to $y.$ Under the usual assumptions, if there are pairs of individuals within and across villages who communicate (receive or transmit information to the others), then everybody in the society communicates and the information transmission relation is transitive. On the other hand, if we assume transitivity, then we can weaken the assumption of the existence of a communication link between any pair of villages with that of a chain of communication links among all the villages.  To be sure, these are fragmented observations that cry out for a systematic treatment.

\section{Concluding Remarks} \lb{sec: conc}

Looked at from far enough, this paper can be easily summarized as revolving around two-and-a-half  theorems:  the first, second and  fourth  can  be combined and collectively presented for  the  case  $k=1,$  as one big portmanteau equivalence theorem   offering a characterization of topological connectedness;\fn{In some sense, we have kept doing versions that approach such a portmanteau theorem.  \label{fn:rg3}}  the third result, combined with the preceeding two for the case $k = 2,$ as a  characterization of 2-connectedness.\fn{The reader is referred to the  counterexample a little above the statement of Theorem 3.} The fifth and sixth results, in giving sufficient conditions for the refinements of transitivity, can be seen as half a theorem of an equivalence that does holds only in one direction.

But one or many, the  six theorems,  and the  propositions that supplement them,  furnish
an overview of a diverse literature in microeconomic theory that is bracketed by a rich mathematical and philosophical literature.\fn{The reader is referred to \cite{te15}, \cite{se17}  and their references;   also to  \cite{an87} and \cite{an93}.   \label{fn:ph}}      Our synthetic treatment of the three remarkable contributions that we sight as foundational, facilitates the reading of past, somewhat neglected, work as well as allows a deeper appreciation of how current contributions
fit into lines of inquiry with a long-established lineage. In showing the sufficiency {\it and} necessity of topological connectedness for both completeness and/or transitivity under one rubric, we generalize and unify these three foundational theorems. To be specific, and perhaps to overly belabour the point, we are not aware of any paper in the literature following Schmeidler that asks for conditions on the topology that follow from his, and  Eilenberg's and Sonnenschein's, behavioral consequences. As such, our results in providing a characterization of topological connectedness for behavior, and therefore for its indispensability, are of interest in and of themselves. We are also not aware of topologizing, and thereby  bringing into a productive relationship, the influential non-topological results of Sen.\fn{As already pointed out in Footnote 3 in the context of the mathematical literature, \cite{wa54} and \cite{bm95} are important references subsequent to Nachbin's pioneering monograph on \lq\lq Topology and Order;" and \cite{mc92} for application to dynamical systems.
The philosophical literature is immense but \cite{an87, an93} furnish an admirable entry into issues concerning the rationality or the irrationality of transitivity and completeness; in addition to the references in \cite{se17} and Temkin, see \cite{tu64}. To be sure, any sharp lines to differentiate between the substantive and the technical eventually leads to sterility.}

Modern decision theory, as charted out by von Neumann, Savage and 
Anscombe-Aumann, is proving fundamental for both behavioural economics  and more generally in issues of empirical inference,\fn{For the first see  \cite{br09} and their references; and for the second, see \cite{pf71} and  \cite{na85}.}  but to the extent that it can be projected to the question  of a  numerical representation of a preference relation over a set of objects, the  ancillary structures both on the set of objects and on the preferences under 
investigation assume a paramount role. 
 The objects could be probability measures, as in von Neumann-Morgenstern; or functions from a state space to a space of consequences, as in Savage; or functions from  a state space to a space of probability measures   on a space of consequences, as in Aumann-Anscombe; or induced relations on the subsets of a space of consequences, as in the menu choices of  Kreps; or  $n$-tuples of preferences over $n$-products of probability measures, as in the temporal lotteries of Kreps-Porteus; but whatever the objects, successful analysis has, of necessity, to involve a play  on the assumptions made on the   preferences and the objects on which these preferences may be defined.\fn{ All this is now so much part of the folklore that detailed references are hardly necessary. But the reader can do no worse than begin with \cite{fi72} and \cite{gi09} on the one hand, and \citet{me98} and \citet{bm95} on the other.  For current activity in the field,  in addition to the three references with which we began Section 4, we refer the reader to \cite{ge17} and \cite{st13} and their references.}  We have limited ourselves solely to the topological register, but already in 1967, Sonnenschein was bringing to bear linear considerations to his deductions concerning transitivity of preferences. To be sure, applications come with a veritable variety of structures, including the linear-algebraic one, with or without a finiteness specification, and how these all interact with each other is a fascinating question that will surely build on the topological treatment explored herein.  What also merits emphasis is that this literature naturally dovetails into work on empirical microeconomic theory, experimental psychology, and philosophical investigation into the very meaning of transitivity, completeness and 
non-satiation, and thereby into specific formalizations of \lq\lq rationality," one of the more vexatious words of our times. We hope to take our results to this theoretical and applied subject matter next.



\section{Proofs of the Results} \lb{sec: proofs}


\setlength{\abovedisplayskip}{4pt}
\setlength{\belowdisplayskip}{3pt}



The presentation of our results in Section \ref{sec: main}  proceeded from the general to the particular: from full equivalence of the $k$-connected case to the 2-connected and connected cases (Theorems \ref{thm: kc} to \ref{thm: cb}), and then finally to the forward (sufficiency) case establishing transitivity in Theorem \ref{thm: ttransitivity}. The motivation for this is simply that it simply gives the readership the option  to read Theorems \ref{thm: kc} and \ref{thm: c} for for the case $k=1,$ and indeed limit themselves only to the forward case. This enables it to see how more restricted settings lead to sharper conclusions. However this is not the best strategy for doing, and presenting, the proofs. It is the more concrete cases that are generalized. As such, we begin with the ancillary Proposition \ref{thm: transitivity} to follow, and use it to prove Theorems \ref{thm: ttransitivity} and \ref{thm: pbtransitivity}. And then Theorem  \ref{thm: ttransitivity} is used in the proof of Theorem  \ref{thm: kc}. These two theorems are used as an input in the proof of Theorem  \ref{thm: c}. The latter is then used to prove Theorems  \ref{thm: 2cb} and \ref{thm: cb}.


\prf[\bf Proof of Proposition \ref{thm: transitivity}]  
We take each claim in turn. 

\smallskip

\nt $\mathbf{\ref{it: tr}}$ The sufficiency of $T$ is due to \citet[Theorem I, assertion I.1]{se69}. In order prove the necessity of $T,$ assume $y\in R(x)$ and $z\in R(y).$ If $y\in R^{-1}(x)$ and $z\in R^{-1}(y),$ then $I \! I$ implies $z\in I(x),$ hence $z\in R(x).$ If $y\notin R^{-1}(x)$, or $z\notin R^{-1}(y),$ or both, then it follows from $P\! P, P\! I, I\! P$ that $z\in P(x),$ hence $z\in R(x).$

\medskip

\noindent  $\mathbf{\ref{it: tnp}}$ Assume $y\in P(x)$ and $z\in P(y).$ It follows from  $y\in P(x)$ and $N\! P$ that  $z\in P(x)\cup  P^{-1}(y)$. Since $z\in P(y)$ and $P$ is asymmetric, therefore   $z\in P(x),$ hence $P\! P$ holds. Now, assume $y\in P(x), z\in I(y)$ and $z\notin P(x).$ It follows from $z\in I(y)$ that $y\notin P(z).$ Then $N\! P$ implies $y\notin P(x).$ This furnishes us a contradiction. Hence, $P\! I$ holds. An analogous argument implies $I\! P.$

\medskip

\noindent $\mathbf{\ref{it: trnp}}$ Let $X$ be a set with at least three elements and $R=\{(x,y)\}$ where $x,y\in X$ and $x\neq y.$ It is clear that $R$ is transitive and $P=R.$ It follows from $(x,z)\notin P, (z,y)\notin P$ and $(x,y)\in P$ for $z\neq x,y$ that $N\! P$ is not satisfied. Now define $R'=\{(x,y), (y,x)\}.$ Then, $P'=\emptyset,$ hence $N\! P$ holds. Since $(x,x)\notin R',$ therefore $T$ is not satisfied. 

\medskip

\noindent  $\mathbf{\ref{it: tindep}}$ We provide a proof by considering examples. Let $x,y,z,w$ be distinct elements of $X.$ We first show $I\! I$ is independent of $P\! P, P\! I, I\! P.$  Define $R=\{(x,y), (y,x), (x,x), (y,y), (y,z), (z,x)\}.$ It is clear that this violates $P\! P, P\! I, I\! P,$ but not $I\!I.$ Next consider $R=\{(x,y), (y,x)\}.$ This violates $I\!I,$  but not $P\! P, P\! I, I\! P.$ 
 Second, we show that $P\!I$ is independent of $P\! P, I\! P,  I\! I.$ Define $R=\{(x,y), (y,z), (x,w), (w,x)\}.$ This violates $P\! P, I\! P, I\! I,$ but not $P\! I.$ Next consider $R=\{(x,y), (y,z), (z,y), (y,y), (z,z)\}.$  This violates $P\! I,$ but not $P\! P, I\! P, I\! I.$ The independence of $I\! P$ is analogously proved and the independence of $P\! P$ is illustrated in \citet[Theorem I, assertion I.2]{se69}. Third, we show that $P\! P, I\! I$ are  independent of $I\! P,  P\! I.$ Define $R=\{(x,y), (y,z), (z,y), (y,y), (z,z), (z,w)\}.$ This violates $I\! P, P\! I,$ but not $P\! P, I\! I.$ Next consider $R=\{(x,y), (y,x), (y,z), (x,z), (z,w)\}.$ This violates $P\! P, I\! I,$ but not $I\! P, P\! I.$ Fourth, we show that $I\! I, P\! I$ are  independent of $P\! P,  I\! P.$ Define $R=\{(x,w), (w,x), (x,x), (w,w), (x,y), (y,z)\}.$ This violates $P\! P, I\! P,$ but not $I\! I, P\! I.$ Next consider $R=\{(x,y), (y,z), (z,y)\}.$ This violates $I\! I, P\! I,$ but not
 $P\! P, I\! P.$ The independence of $I\! I, I\! P$ and $P\! P,  P\! I$ can be shown analogously. 
\medskip

The proof of Proposition \ref{thm: transitivity} is complete. 
\prff


Next we turn to the proof of Theorem \ref{thm: ttransitivity}. Before that we need the following definition and a lemma.

\df
A partition of a set $X$ is a collection of non-empty and pairwise disjoint sets $\{A_\lambda\subset X~|~ \lambda\in \Lambda\}$ such that $\bigcup_{\lambda\in \Lambda} A_\lambda=X.$ A partition is open if all of its members are open, and a partition is closed if all of its members are closed. 
\lb{df: par}
\dff

\lm
For all  semi-transitive and continuous binary relation $R$  on a topological space $X$,  all components $C, C'$ of $X$ and all $x\in C, y\in C'$, if $(x,y)\in P,$ then $C\cup C' \subset P(x)\cup P^{-1}(y)$.
\lb{thm: ccover}
\lmm

\prf [Proof of Lemma \ref{thm: ccover}.]  
 Pick, possibly identical, two components $C,C'$ of $X$ and $x\in C, y\in C'$ such that $(x,y)\in P$. Then, $x\in P^{-1}(y)$ and  $y\in P(x).$ Hence $P(x)\cup P^{-1}(y)$ has non-empty intersections with both $C$ and $C'.$  It follows from $P$ has open sections that $P(x)\cup P^{-1}(y)$ is open.  Since $R$ has closed sections, therefore  $R(x)\cup R^{-1}(y)$ is closed.  If $P(x)\cup P^{-1}(y) = R(x)\cup R^{-1}(y),$ then we have a subset of $X$ which is both open and closed, and has non-empty intersection with  both $C$ and $C'.$ Since $C$ and $C'$ are components of $X,$ therefore    $C\cup C'\subset P(x)\cup P^{-1}(y).$ It remains to prove  $P(x)\cup P^{-1}(y) = R(x)\cup R^{-1}(y).$ 
 
  It is clear that $P(x)\cup P^{-1}(y) \subset R(x)\cup R^{-1}(y).$ In order to show  the reverse inclusion assume there exists $z\in R(x)\cup R^{-1}(y)$ such that $z\notin P(x)$ and $z\notin P^{-1}(y).$
 If  $z\in R(x),$ then  it follows from $z\notin P(x)$ that $x\in R(z).$ Hence $(z,x)\in I.$ It follows from $I\! P$ and $(x,y)\in P$ that $z\in P^{-1}(y).$ This furnishes us a contradiction. 
   If $z\in R^{-1}(y),$ then it follows from $z\notin P^{-1}(y)$ that $z\in R(y).$ Hence $(y,z)\in I.$ It follows from $P\! I$ and $(x,y)\in P$ that $z\in P(x).$ This furnishes us a contradiction.   Therefore $R(x)\cup R^{-1}(y) \subset P(x)\cup P^{-1}(y)$. 
\prff


\prf[\bf Proof of Theorem \ref{thm: ttransitivity}]  We assume that $R$ is a continuous binary relation on a topological space $X$ and begin the proof of each claim in $\mathbf{\ref{it: tippi}}$ under the assumption that the topology on $X$ is connected. 
\smallskip



\nt $\mathbf{\ref{it: tippi1}}$ Note that $P$ is negatively transitive if and only if for all $x,y,z\in X$, $(x,y)\in P$ implies either $(x,z)\in P$ or $(z,y)\in P$.  Pick $x,y\in X$ such that $(x,y)\in P.$  Since $X$ is connected, it follows from Lemma \ref{thm: ccover} that  $X\subset P(x)\cup P^{-1}(y)$. Hence, $P$ is negatively transitive. 
 The backward direction follows from the assertion \ref{it: tnp} in Proposition \ref{thm: transitivity}.
\medskip
   
\nt $\mathbf{\ref{it: tippi2}, \ref{it: tippi3}, \ref{it: tippi4}}$ The proofs follow from the assertions \ref{it: tippi1} above, and  \ref{it: tr}, \ref{it: tnp} in Proposition \ref{thm: transitivity}. 
\medskip


Next, we turn to the proof of each claim in  $\mathbf{\ref{it: tii}}$ under the assumption that the sections of $I$ are connected.
\smallskip


\nt $\mathbf{\ref{it: tii1}}$ Pick $x,y,z\in X$ such that $y\in P(x)$ and $z\in I(y).$ Assume $z\notin P(x).$ 
Then, it follows from  $I\! I$ that $I(x)\cap I(z)=\emptyset$.  Then  $X=I(x)\cup P(x)\cup P^{-1}(x) \cup \left(R(x)\cup R^{-1}(x)\right)^c$ implies that  
$$
I(z)=\left[ P(x) \cap I(z)\right]  \cup \left[ P^{-1}(x)\cap I(z) \right] \cup \left[   (R(x))^c\cap \left(R^{-1}(x)\right)^c \cap I(z) \right].  
$$
It is clear that the three sets in square brackets are pairwise disjoint. Since $P$ has open sections and $R$ has closed sections, the three sets in square brackets are open in $I(z).$ Since $(z,y)\in I$ and $I$ is symmetric, therefore $(y,z)\in I$. Then $I\! I$ implies $(z,z)\in I$. Therefore, $y,z\in I(z).$ 
  It is clear that $y\in P(x)\cap I(z)$.  Since we assume above that $z\notin P(x)$, therefore $z$ is either in $P^{-1}(x)$ or in $(R(x))^c\cap \left(R^{-1}(x)\right)^c$, but not in both since these two sets are disjoint. 
If $z\in P^{-1}(x),$ then $P(x) \cap I(z)$ and the union of the remaining two sets in square brackets above form an open partition of $I(z)$ which contradicts the connectedness of $I(z).$ Analogously, $z\in (R(x))^c\cap \left(R^{-1}(x)\right)^c$ furnishes us a contradiction to the connectedness of $I(z).$ Therefore, $z\in P(x),$ and hence $P\! I$ holds. An analogous argument implies $I\! P$ holds. 
\medskip

\nt $\mathbf{\ref{it: tii2}}$ The proof follows from assertions \ref{it: tii1} above and \ref{it: tr}  in Proposition \ref{thm: transitivity}.

\medskip

The proof of Theorem \ref{thm: ttransitivity} is complete. 
\prff


We now turn to the proof of Theorem \ref{thm: pbtransitivity}.

\prf[{\bf Proof of Theorem \ref{thm: pbtransitivity}}]

Let $X$ be a topological space with the Phragmen-Brouwer property and $R$ a continuous relation on it with path-connected sections. We take each claim in turn. 
\medskip

\nt $\mathbf{\ref{it: tpb1}}$ Assume the section of the symmetric part $I$ of $R$ is transitive. Now consider the following claim. 
 \cl 
 For all $z\in X,$ the sets $P(z)$ and $P^{-1}(z)\cup \left(R(z)\cup R^{-1}(z)\right)^c$ are separated. 
 \lb{thm: separated}
 \cll

 Pick $x,y,z\in X$ such that $x\in I(y)$ and $y\in P(z).$ Assume $x\notin P(z).$ If $x\in I(z),$ then it follows from the transitivity of $I$ that $y\in I(z),$ which contradicts with $ y\in P(z).$ Therefore, either $x\in P^{-1}(z)$ or $x\in \left(R(z)\cup R^{-1}(z)\right)^c.$  
   Recall that  $ x\in P^{-1}(z)\cup \left(R(z)\cup R^{-1}(z)\right)^c, y\in P(z)$ and 
   $$
   X\backslash I(z)=P(z)\cup P^{-1}(z)\cup \left(R(z)\cup R^{-1}(z)\right)^c.
   $$
 Claim \ref{thm: separated} implies that the sets $P(z)$ and  $P^{-1}(z)\cup \left(R(z)\cup R^{-1}(z)\right)^c$ are separated. Then it follows from $X$ has the Phragmen-Brouwer property that there exists a connected subset\fn{Note that we can choose this connected subset as the component $C$ of $I(z)$ which contains $I_z$. This follows from the fact that if a set is connected in a subspace $Y$ of a topological space, then it is also connected in any subspace containing $Y$. In order to see this, note that it follows from $I_z$ separates $x$ and $y$ that there exist distinct components $C_x$ and $C_y$ of $X\backslash I_z$ such that $x\in C_x$ and $y\in C_y$. Assume that $x$ and $y$ are contained in the same component $C_{xy}$ of $X\backslash C.$  Since $C$ contains $I_z,$ therefore $X\backslash C\subset X\backslash I_z.$ Hence $C_{xy}$ is connected in $X\backslash I_z.$ Then $x\in C_x\cap C_{xy}$ and $y\in C_y\cap C_{xy}$ contradict $C_x$ and $C_y$ being distinct components of $X\backslash I_z.$ Therefore,  $x$ and $y$ are contained in distinct components of $X\backslash C,$ i.e., $C$ separates $x$ and $y.$ Since any component of a space is connected, the requirement that the subset $I_z$ is closed in the definition of the Phragmen-Brouwer property in \citet[ Property V$'$, p. 50]{wi49} is not restrictive. However, our version requires the separation property holds {\it only} for open sets.}  
 $I_z$ of $I(z)$ which  separates $x$ and $y.$ 
 Note that $I\! I$ and $(x,y)\in I$ imply that $x,y\in R(y)$. Since $R(y)$ is path-connected, therefore there exists a continuous function $s:[0,1]\ra R(y)$ such that $s(0)=x$ and  $s(1)=y.$ Note that $s([0,1])$ is connected and $x,y$ are contained in different components of $X\backslash I_z,$ therefore there exists  $\lambda\in [0,1]$ such that $s(\lambda)=z'\in I_z\cap R(y).$ Since $I_z\subset I(z)$, it follows from $I\! I$ and $y\in P(z)$ that $I_z\cap I(y)=\emptyset.$ Hence $z'\in P(y)$ and 
 $$
I_z=\left[ P(y) \cap I_z\right]\cup \left[ P^{-1}(y) \cap I_z\right]\cup \left[ \left(R(y)\cup R^{-1}(y)\right)^c \cap I_z\right].
 $$
 Since $I_z$ is connected and $z'\in P(y),$ therefore $I_z\subset P(y).$ 
 
 Claim \ref{thm: separated} implies that $P(y)$ and $P^{-1}(y)\cup \left(R(y)\cup R^{-1}(y)\right)^c$ are separated subsets of $X$. Then, it follows from $z'\in P(y)$ and $z\in  P^{-1}(y)$ that  there exists a connected subset $I_y$ of $I(y)$ which separates $z'$ and $z.$ Note that $z'\in I_z\subset I(z)$ and $I\! I$ imply that $(z,z)\in I$.  Then  $z,z'\in R(z')$ and $R(z')$ is path-connected imply that there exists a continuous function $s':[0,1]\ra R(z')$ such that $s'(0)=z$ and  $s'(1)=z'.$ Note that $s'([0,1])$ is connected and $z,z'$ are contained in different components of $X\backslash I_y,$ therefore there exists $\lambda\in [0,1]$ such that $s'(\lambda)=y'\in I_y.$  It follows from the transitivity of $I$ that $I_y\cap I(z')=\emptyset,$ therefore 
 $$
I_y=\left[ P(z') \cap I_y\right]\cup \left[ P^{-1}(z') \cap I_y\right]\cup \left[ \left(R(z')\cup R^{-1}(z')\right)^c \cap I_y\right].
 $$  
 Since $I_y$ is connected and $y'\in P(z'),$ therefore $I_y\subset P(z').$ 
 
 It follows from $z'\in P^{-1}(y')$ and $I_z$ is connected that $I_z\subset P^{-1}(y').$ Therefore, $R(y')\subset X\backslash I_z.$ Then $x,y\in R(y')$ and $x,y$ are contained in distinct components of $X\backslash I_z$ contradicts with $R(y')$ is connected. Therefore, $x\in  P(z),$ hence $P\! I$ holds.

 It remains to prove Claim \ref{thm: separated} in order to finish the proof of \ref{it: tpb1}.


 \prf[Proof of Claim \ref{thm: separated}]
 Pick $z\in X$ and define $A=P^{-1}(z)\cup \left(R(z)\cup R^{-1}(z)\right)^c$.  Since $R(z)$ is closed, therefore $\overline{P(z)}\subset R(z)$.  It is clear that $R(z)\cap A=\emptyset$, hence  $\overline{P(z)}\cap A=\emptyset$. It remains to show that $\overline A\cap P(z)=\emptyset$. It follows from  $P(z)$ is open, $A\cup I(z)\cup P(z)=X$ and $(A\cup I(z))\cap P(z)=\emptyset$ that $A\cup I(z)=\left(P(z)\right)^c$ and $A\cup I(z)$ is closed. Therefore $\bar A\subset A\cup I(z)$, and hence $\bar A\cap P(z)=\emptyset$.  Therefore, $P(z)$ and $P^{-1}(z)\cup \left(R(z)\cup R^{-1}(z)\right)^c$ are separated sets in $X$. 
 \prff

\nt $\mathbf{\ref{it: tpb2}}$ Replacing $R$ and $P$ with $R^{-1}$ and  $P^{-1}$ in the proof of claim \ref{it: tpb1} above completes the proof. 
\medskip

\nt $\mathbf{\ref{it: tpb3}}$ The proof follows from claims \ref{it: tpb1} and  \ref{it: tpb2}  above and Proposition \ref{thm: transitivity}\ref{it: tr}.
\medskip

The proof of Theorem \ref{thm: pbtransitivity} is complete. 
\prff

We now turn to the proof of Theorem \ref{thm: kc}. Before that we need the following two lemmata. 

\lm
Any non-empty, closed and open subset of a topological space is a union of the components of the space.
\lb{thm: components}
\lmm

\prf[\bf Proof of Lemma \ref{thm: components}.]   
   Pick a non-empty, closed and open subset $V$ of a topological space $X.$ It follows from \citet[Theorem 3.2]{du66} that the components of $X$ form a closed partition of $X.$ Let $C_x$ denote the component of $X$ containing $x\in X.$ We next show that 
 $
V=\bigcup_{x\in V}C_x.
$  
It is clear that $V\subset\bigcup_{x\in V}C_x.$ In order to show the reverse inclusion, pick $x\in V,$ define  $A_x=C_x\cap V$ and $B_x=C_x\cap V^c.$ Since $C_x, V, V^c$ are closed, therefore $A_x, B_x$ are closed. Moreover, $A_x\cup B_x=C_x$ and $A_x\cap B_x=\emptyset.$ Then $A_x=B_x^c\cap C_x$ and $B_x=A_x^c\cap C_x.$ Hence, $A_x$ and $B_x$ are both open and closed in the subspace $C_x.$  Since $C_x$ is connected, therefore either $A_x=\emptyset$ or $B_x=\emptyset.$ It follows from $x\in A$ that $B_x=\emptyset.$ Then $A_x=C_x.$ Since $A_x=C_x\cap V,$ therefore $C_x\subset V.$
\prff

\lm
Any topological space with at least  $k$ components has a partition consisting of $k$ sets which are both open and closed. 
\lb{thm: partition}
\lmm

\prf[\bf Proof of Lemma \ref{thm: partition}.]   
 Pick a natural number $k$ and assume  $X$ is a topological space with at least $k$ components. If $k=1$, then setting $\{X\}$ as the partition of $X$ completes the proof. Otherwise, $X$ is disconnected, hence there exist $A_1, A_1^c$ which form a partition of $X$ which is both open and closed. If $k=2$,  then the proof is complete.  
Otherwise, Lemma \ref{thm: components} implies that $A_1$ and $A_1^c$ can be written as a union of the components of $X.$ Since the components of $X$ are disjoint, therefore $A_1$ and $A_1^c$ are the unions of distinct  components. Let $A_1^c$ be the union of at least two components. Then, $A_1^c$ is disconnected, hence there exist non-empty subsets $A_2, A_2^c$ of $A_1^c$ which are both open and closed in the subspace $A_1^c.$ Since $A_1^c$ is open and closed in $X$, therefore $A_2, A_2^c$ are also open and closed in $X.$ Then, $A_1, A_2, A_2^c$ form a partition of $X$ which is both open and closed.  If $k=3$, then the proof is complete. Otherwise, repeating this procedure $(k-1)$-many times yields an open partition of $X$ consisting of $k$ sets.
\prff




\prf[\bf Proof of Theorem \ref{thm: kc}.]  
We begin the proof with $\mathbf{\ref{it: kc} \Rightarrow \ref{it: kess}}$. The proof rests on three claims which we state and use, and prove only after the proof of the implication \ref{it: kc} $\Rightarrow$ \ref{it: kess} is complete. 
 Assume $X$ is $k$-connected and $\{C_1, \ldots, C_k\}$ denote the set of components of $X.$ Define $K=\{1,\ldots, k\}.$  Let $R$ be a $k$-non-trivial, semi-transitive and continuous binary relation on $X$ such that its symmetric part is transitive.  
 %
 %
  The following claim shows that every pair of components contains strictly comparable elements.
\cl
For all $i,j\in K,$ there exists $x_i\in C_i, x_j\in C_j$ such that $(x_i, x_j)\in P\cup P^{-1}.$  
\lb{thm: comp1}
\cll

Assume there exist $x,y\in X$ such that $(x,y)\notin R\cup R^{-1}.$ Then $x\in C_i$ and $y\in C_j$ for some $i,j\in K$.  Claim \ref{thm: comp1} implies that there exist $x_i\in C_i, x_j\in C_j$ such that $(x_i, x_j)\in P\cup P^{-1}.$ Without loss of generality, assume $(x_i, x_j)\in P.$ Then, it follows from Lemma \ref{thm: ccover} that $x,y\in P(x_i)\cup P^{-1}(x_j).$ The following claim shows that both $x$ and $y$ are contained in at least one of these two sets. 

\cl
$\{x,y\}\subset P(x_i)$ or $\{x,y\}\subset P^{-1}(x_j)$.
\lb{thm: comp2}
\cll

\noindent It follows from  Claim \ref{thm: comp2} that $x_i\in P^{-1}(x)\cap P^{-1}(y)$ or  $x_j\in P(x)\cap P(y).$ Therefore,  $\left[P^{-1}(x)\cap P^{-1}(y)\right]\cap C_i\neq \emptyset$ or $\left[P(x)\cap P(y)\right]\cap C_j\neq \emptyset.$ It follows from  $x\in C_i, y\in C_j,$  $x, y\notin P^{-1}(x)\cap P^{-1}(y)$ and $x, y\notin P(x)\cap P(y)$ that $C_i, C_j\not\subset P^{-1}(x)\cap P^{-1}(y)$ and $C_i, C_j\not\subset P(x)\cap P(y).$

\cl
$P^{-1}(x)\cap P^{-1}(y)=R^{-1}(x)\cap R^{-1}(y)$ and $P(x)\cap P(y)=R(x)\cap R(y)$. 
\lb{thm: comp3}
\cll

\noindent  It follows from continuity of $R$ and Claim \ref{thm: comp3} that  $P(x)\cap P(y)$ and $P^{-1}(x)\cap P^{-1}(y)$ are both open and closed. Then $\left\{P^{-1}(x)\cap P^{-1}(y)\cap C_i, \left[P^{-1}(x)\cap P^{-1}(y)\right]^c\cap C_i\right\}$ is an open partition of $C_i$ or $\left\{P(x)\cap P(y)\cap C_j, \left[P(x)\cap P(y)\right]^c\cap C_j\right\}$ is an open partition of $C_j.$  This furnishes us a contradiction to $C_i$ and $C_j$ being components of $X.$ Therefore, $R$ is complete and hence the proof of assertion \ref{it: kc} $\Rightarrow$ \ref{it: kess} is complete. 

It now remains to prove Claims \ref{thm: comp1} -- \ref{thm: comp3}.
\vspace{-5pt}

\prf[Proof of Claim \ref{thm: comp1}]
First, pick $i,j\in K$ such that $i=j$. It follows from $k$-non-triviality that there exist $\bar x_j, \bar y_j\in C_j$ such that $(\bar x_j,\bar y_j)\in P.$ Then Lemma \ref{thm: ccover} implies that $C_j\subset P(\bar x_j)\cup P^{-1}(\bar y_j)$.  Second, pick $i,j\in K$ such that $i\neq j.$ Then $k$-non-triviality implies  that there exist $x_i\in C_i,  x_j\in C_j$ such that $(x_i,  x_j)\in R\cup R^{-1}.$ Assume without loss of generality that $(x_i, x_j)\in R.$ If $(x_i, x_j)\in P,$ then the proof is complete. Then, let $(x_i, x_j)\in I.$ Since $x_j\in C_j\subset P(\bar x_j)\cup P^{-1}(\bar y_j)$, therefore the semi-transitivity of $R$ implies that $(x_i,  \bar y_j)\in P$ or $(x_i, \bar x_j)\in P^{-1}$.
\prff

\prf[Proof of Claim \ref{thm: comp2}.] 
  If $x\in P^{-1}(x_j),$ then Lemma \ref{thm: ccover} implies that $y\in P(x)\cup P^{-1}(x_j).$ Since $y\notin P(x),$ therefore $y\in P^{-1}(x_j).$ 
  If $x\in P(x_i),$ then it follows from $x,x_i\in C_i$ and Lemma \ref{thm: ccover} that $P(x_i)\cup P^{-1}(x)$  is both open and closed and contains $C_i.$ Since $x_j\in P(x_i)\cap C_j,$ therefore  $P(x_i)\cup P^{-1}(x)$ has a non-empty intersection with $C_j.$ Then, it follows from $P(x_i)\cup P^{-1}(x)$  is both open and closed, and $C_j$ is a component of $X$ that $C_i\cup C_j\subset P(x_i)\cup P^{-1}(x).$ Hence, $y\in P(x_i)\cup P^{-1}(x).$ Since $y\notin P^{-1}(x),$ therefore $y\in P(x_i).$ 
\prff

\prf[Proof of Claim \ref{thm: comp3}.]  
It is clear that $P(x)\cap P(y) \subset R(x)\cap R(y).$   In order to show  the reverse inclusion  assume there exists $z\in R(x)\cap R(y)$ such that $z\notin P(x)$ or $z\notin P(y).$ If $z\notin P(x),$ then it follows from $z\in R(x)$ that $(z,x)\in I.$ Then, $z\in R(y),$ $I\! I$ and $P\! I$ imply either $(y,x)\in I$ or $(y,x)\in P.$ This furnishes us a contradiction to $(x,y)\notin R\cup R^{-1}.$   
 If $z\notin P(y),$ then it follows from $z\in R(y)$ that $(z,y)\in I.$ Then, $z\in R(x),$ $I\! I$ and $I\! P$ imply either $(x,y)\in I$ or $(x,y)\in P.$ This furnishes us a contradiction to $(x,y)\notin R\cup R^{-1}.$ Therefore, $R(x)\cap R(y) \subset P(x)\cap P(y)$.
 
 Replacing $R$ with $R^{-1}$ and $P$ with $P^{-1}$ in the argument above implies that $P^{-1}(x)\cap P^{-1}(y)=R^{-1}(x)\cap R^{-1}(y)$. 
\prff

Next we turn to the other implications in Theorem \ref{thm: kc}. 
\smallskip

\noindent $\mathbf{\ref{it: kc} \Rightarrow \ref{it: kes}}$ Theorem \ref{thm: ttransitivity}\ref{it: tii} implies $R$ is semi-transitive.  Then $\text{\ref{it: kc}} \Rightarrow \text{\ref{it: kess}}$ above completes the proof. 

\medskip

\noindent $\mathbf{\ref{it: kc} \Rightarrow \ref{it: keg}}$  The proof follows from \ref{it: kc} $\Rightarrow$ \ref{it: kess} above and the observation that the anti-symmetry of $R$ implies that $R$ is semi-transitive and $I$ is transitive. In order to see this note that it follows from the  anti-symmetry of $R$ that $I(x)\subset \{x\}.$ Then, if $y\in P(x)$ and $z\in I(y),$ then $z=y,$ hence $z\in P(x).$ If $y\in I(x)$ and $z\in P(y),$ then $x=y,$ hence $z\in P(x).$ Similarly, if $y\in I(x)$ and $z\in I(y),$ then $z=y=x,$ hence $z\in I(x).$ 

\medskip

\noindent $\mathbf{\ref{it: kc} \Rightarrow \ref{it: ksc}}$ The proof follows from \ref{it: kc} $\Rightarrow$ \ref{it: kess} since Proposition \ref{thm: transitivity}\ref{it: tr} implies that $R$ is semi-transitive and its symmetric part is transitive.

\medskip

\noindent $\mathbf{\ref{it: kess}, \ref{it: kes}, \ref{it: keg}, \ref{it: ksc} \Rightarrow \ref{it: kc}}$   Assume  $X$ has at least $k+1$ components. Then Lemma \ref{thm: partition} implies that there exists a partition $
\{Y_1, Y_2\ldots, Y_{k+1}\}$ of $X$ which is both open and closed.  
 %
%
  Define a binary relation $R$ on $X$ as 
$$
R=\bigcup_{i=1}^{k} \left(\bigcup_{j=i+1}^{k+1} Y_i \times Y_j\right).
$$ 
  Then the symmetric part of $R$ is $I=\emptyset$ and its asymmetric part is $P=R.$ By construction, the sections of $R$ are closed and the sections of $P$ are open. Moreover, $R$ is transitive, semi-transitive and anti-symmetric, and $I$ is transitive. Defining $m_i=i$ and $n_i=i+1$ for all $i\leq k+1$ imply that $R$ is $k$-non-trivial. Finally, it is clear that $R$ is incomplete.    
\smallskip

 The proof of Theorem \ref{thm: kc} is complete.
\prff


Before turning  to the proof of Theorem \ref{thm: c}, we comment on our proof-technique. Eilenberg uses completeness of an anti-symmetric and continuous relation on a connected space in order to obtain transitivity.  Sonnenschein exploits his standard quotient-space construction  to drop the anti-symmetry assumption in Eilenberg's theorem. He bring into prominence the assumption of semi-transitivity of the relation which is satisfied by any anti-symmetric relation, and then provides a series of sufficient conditions for semi-transitivity by imposing further topological assumptions on preferences. On the other hand Schmeidler uses transitivity of a non-trivial and continuous relation on a connected space in order to obtain completeness. Although both use the connectedness of the space and the continuity of the relation, the proof techniques of Schmeidler and of Eilenberg-Sonnenschein are quite different. In the proofs of Theorems \ref{thm: kc} and \ref{thm: ttransitivity} above, the latter has transitivity and the former, completeness as its necessary condition. Our proof-technique  is inspired by that of Schmeidler, and we use it to obtain independently each of the completeness {\it and} the transitivity properties
in  the forward-direction of Theorem \ref{thm: c}.  
This alternative proof of the results of Eilenberg and Sonnenschein may have some independent interest.
  
\medskip   
  
 We are now ready to prove Theorem \ref{thm: c}.

\prf[\bf Proof of Theorem \ref{thm: c}.] 

We first prove the forward direction  and then turn to the proof of the backward direction.  In the forward direction, for each assertion, completeness follows from its counterpart in Theorem \ref{thm: kc}. Hence, the proof of the implication \ref{it: c}$\Rightarrow$\ref{it: sc} is complete. We next prove that transitivity holds in the remaining implications without using the completeness property.  If $X$ has only one component, then transitivity follows from Proposition \ref{thm: transitivity} and Theorem \ref{thm: ttransitivity}. Hence, assume $X$ has two components $C_1, C_2$ which form a partition of $X$ that is both open and closed. 
\medskip

\nt $\mathbf{\ref{it: c} \Rightarrow \ref{it: ess}}$. Assume $R$ is a $2$-non-trivial, semi-transitive and continuous binary relation on $X$ with a transitive symmetric part.
 %
Pick $x,y,z\in X$ such that $y\in R(x)$ and $z\in R(y).$  If $x\in R(y)$ or $y\in R(z)$ holds, then the semi-transitivity of $R$ and  the transitivity of $I$ imply $z\in R(x)$. Hence, assume  $y\in P(x)$ and $z\in P(y).$  Since $C_1$ and $C_2$ form a partition of $X,$ each of $x,y,z$ is contained in one and only one of these two components. Then the following four cases cover all possibilities: (i) $x,y,z \in C_i,$ (ii) $x, y\in C_i, z\in C_j$, (iii) $x, z\in C_i, y\in C_j$ and (iv) $x\in C_i, y,z\in C_j$ where $i=1,2, i\neq j.$

Before elaborating the cases note that  it follows from Lemma \ref{thm: ccover}  that for all $i,j=1,2$, if there exist $x_i\in C_i$ and $y_j\in C_j$ such that $(x_i,y_j) \in P$,  then $C_i\cup C_j\subset P(x_i)\cup P^{-1}(x_j)$. 
Then in cases (i), (ii) and (iii), Lemma \ref{thm: ccover}  implies that $x\in P(y)\cup P^{-1}(z)$. Then it follows from $y\in P(x)$ that $z\in P(x)$.  
Similarly, in case (iv), Lemma \ref{thm: ccover}  implies that $z\in P(x)\cup P^{-1}(y)$. Then it follows from $z\in P(y)$ that $z\in P(x)$.  Therefore, $R$ is transitive. 
\medskip
 
\nt $\mathbf{\ref{it: c} \Rightarrow \ref{it: es}}$  The proof follows from \ref{it: c} $\Rightarrow$ \ref{it: ess} above and Theorem \ref{thm: ttransitivity}\ref{it: tii}. 
\medskip

\nt $\mathbf{\ref{it: c} \Rightarrow \ref{it: eg}}$ The proof follows from \ref{it: c} $\Rightarrow$ \ref{it: ess} above and the observation that any anti-symmetric relation is semi-transitive and its symmetric part is transitive. 

\medskip

The proof of the forward direction is complete. We provide the proof of the backward direction by considering cases $k=1$ and $k=2$ separately. 
\medskip

 We begin with the case $k=1$.  
\smallskip

\nt  $\mathbf{\ref{it: ess}, \ref{it: es}, \ref{it: eg}, \ref{it: sc} \Rightarrow \ref{it: c}}$ Assume $X$ is disconnected. Then there exists an open partition of $X$ consisting of a set $Y$ and its complement $Y^c.$ Define $R= Y\times Y^c.$  Then  it is asymmetric, hence  $I=\emptyset$ and   $P=R.$ Since $Y$ and $Y^c$ are both open and closed, therefore $R(x)=Y^c=P(x)$ and $R^{-1}(x)=Y=P^{-1}(x)$ are both open and closed for all $x\in X.$ Since $P\neq \emptyset,$ $R$ is non-trivial. It follows from $P=R$ that $R$ is anti-symmetric.  It is clear that $R$ is transitive, hence semi-transitive. Moreover, since $I=\emptyset,$ therefore it is transitive and its sections are connected. Since $Y$ and $Y^c$ are non-empty,  therefore $R\cup R^{-1}\neq X\times X,$ i.e. $R$ is not complete.  This furnishes us a contradiction. 
\medskip


 Next we turn to the case $k=2$.  

%
%
\smallskip
 
\nt $\mathbf{\ref{it: ess}, \ref{it: es}, \ref{it: eg}, \ref{it: sc} \Rightarrow \ref{it: c}}$   Assume  $X$ has at least three components. Then Lemma \ref{thm: partition} implies that there exists a partition  $\{Y_1, Y_2, Y_3\}$ of $X$ which is both  open and closed. Define a binary relation on $X$ as $R=\left(Y_1\times Y_2\right) \cup\left( Y_1\times Y_3\right) \cup \left(Y_2\times Y_3\right).$  Then it is asymmetric, hence  $I=\emptyset$ and   $P=R.$ By construction, the sections of $R$ is closed, of $P$ are open and of $I$ are connected. Moreover, $R$ is transitive, semi-transitive and anti-symmetric, and $I$ is transitive. Defining $\mathcal C^1=\{Y_1, Y_2\}$ and $\mathcal C^2=\{Y_2, Y_3\}$ implies $R$ is $2$-non-trivial. Finally, it is clear that $R$ is incomplete.    
\medskip

 The proof of Theorem \ref{thm: c} is complete.
\prff

\medskip

\nt {\bf Remark:} We could have also relied on  the proof-technique of Eilenberg-Sonnenschein in order to prove Theorem \ref{thm: c}.  This would require  reflexivity of the preference relation. Since the assumptions of the theorem imply the completeness of the relation, and hence its reflexivity, this is not a restrictive assumption. Moreover, this method requires the following intermediate result on the relation between the number of the components of a space and of its quotient space, and whose proof is an easy consequence of Lemma \ref{thm: partition}. 
\medskip

\nt {\it  Let $X$ be a topological space, $I$ an equivalence relation on it and $X|I$ the quotient space of it with respect to $I.$  If $X$ has $k$ components, then $X|I$ has at most $k$ components. Moreover, if $X|I$ has $k$ components, then $X$ has at least $k$ components.  }
\medskip

\nt   If the section of $I$ are connected, then this result suggests that the connectedness of a space $X$ is equivalent to the connectedness of its quotient $X| I$. Therefore,  the weaker connectedness assumption of \citet[Theorem 4]{so65} ($X| I$ is connected) is equivalent to the stronger connectedness assumption ($X$ is connected). 

\medskip


We now turn to the proof of Theorem \ref{thm: 2cb}.

\prf[\bf Proof of Theorem \ref{thm: 2cb}.]  
In the forward direction, each assertion is a special case of its counterpart in Theorem \ref{thm: c}, hence the proof of this direction is complete. Note that we have not used the assumption that the space is quasi-ordered yet. 
\medskip

\noindent $\mathbf{\ref{it: son}, \ref{it: son2}, \ref{it: eil} \Rightarrow \ref{it: 2cb}}$  Assume  $X$ is a quasi-ordered space and has at least three components. Then Lemma \ref{thm: partition} implies that there exists a partition $
\{Y_1, Y_2, Y_{3}\}$ of $X$ which is both open and closed.   
Let $Q$ be a complete, anti-symmetric and continuous binary relation (since the space is quasi-ordered, such relation exists). Define a binary relation $R$ on $X$ as follows. 
\[
\begin{array}{ll}
\forall x,y\in Y_i,  & x\in R(y) \Leftrightarrow x\in Q(y),~ i=1,2,3,\\ 
\forall x\in Y_1, \forall y\in Y_2,  & y\in R(x),\\
\forall y\in Y_2, \forall z\in Y_3,  & z\in R(y),\\ 
\forall x\in Y_1, \forall z\in Y_3,  & x\in R(z).\\
\end{array}
\] 
We next show that $R$ is complete and anti-symmetric with closed sections, but non-transitive. It follows from $Q$ is complete and anti-symmetric that $R$ is complete and anti-symmetric. Pick $x\in Y_1, y\in Y_2, z\in Y_3.$ Then
 \[
\begin{array}{c}
R(x)=\left(  Q(x)\cap Y_1\right) \cup Y_2 ~\mbox{ and }~ R^{-1}(x) =\left(  Q^{-1}(x)\cap Y_1\right) \cup Y_3, \\
 R(y) =\left(  Q(y)\cap Y_2\right) \cup Y_3 ~\mbox{ and }~ R^{-1}(y) =\left(  Q^{-1}(y)\cap Y_2\right) \cup Y_1, \\
R(z) =\left(  Q(z)\cap Y_3\right) \cup Y_1 ~\mbox{ and }~  R^{-1}(z) =\left(  Q^{-1}(z)\cap Y_3\right) \cup Y_2. \\
\end{array}
\] 
It follows from $Q$ has closed sections and $Y_1, Y_2, Y_3$ are closed that $R$ has closed sections. Since $x\in Y_1, y\in Y_2, z\in Y_3$ implies $y\in R(x), z\in R(y)$ and $x\in R(z),$ the relation $R$ is non-transitive. This furnishes us a contradiction. Hence $X$ is 2-connected.
\smallskip

The proof of Theorem \ref{thm: 2cb} is complete.
\prff



We now turn to the proof of Theorem \ref{thm: cb}. 

\prf[\bf Proof of Theorem \ref{thm: cb}.]      

\noindent $\mathbf{\ref{it: c2} \Rightarrow \ref{it: sccont}}$ Assume $R$ is incomplete. Let $P$ denote its asymmetric part. Note that it follows from the negative transitivity of $P$ that
\eq
\mbox{if $(x,y)\in P,$ then $P(x)\cup P^{-1}(y)=X.$}
\lb{eq: strict0}
\eqq

\noindent Since $R$ is incomplete, there exists $x,y\in X$ such that $(x,y)\notin R\cup R^{-1}.$ Then  $P^{-1}(x)\cap P^{-1}(y)$ and $P(x)\cap P(y)$ are proper subsets of $X.$  Since $R$ is strongly non-trivial, there exist $(\bar x, \bar y)\in P$ such that $R(x')\cap R(y')\neq \emptyset$ for all $x',y'\in P(\bar x).$  It follows from (\ref{eq: strict0}) above that $x\in P(\bar x)\cup P^{-1}(\bar y).$ 
 If $x\in P(\bar x),$ then (\ref{eq: strict0}) implies that $y\in P(\bar x)\cup P^{-1}(x).$ Since $y\notin P^{-1}(x)$, therefore $y\in P(\bar x).$ 
  Since $x, y\in P(\bar x),$ there exists $z\in X$ such that  $z\in R(x)\cap R(y).$    
   If $x\in P^{-1}(\bar y),$ then (\ref{eq: strict0}) implies that $y\in P(x)\cup P^{-1}(\bar y).$ Since $y\notin P(x)$, $y\in P^{-1}(\bar y).$Hence, $\bar y\in P(x)\cap P(y)$. Therefore, we established that  $R(x)\cap R(y)\neq \emptyset.$

Since $R$ is semi-transitive and $I$ is transitive, therefore Claim \ref{thm: comp3} in the proof of Theorem \ref{thm: kc} implies that $R(x)\cap R(y)=P(x)\cap P(y).$  Then there exists  a non-empty and proper subset of $X$ which is both open and closed. This furnishes us a contradiction to the connectedness of $X.$ Therefore, $R$ is complete. Then, the completeness of $R$ and the negative transitivity of $P$ implies the transitivity of $R$.  Since $R^{-1}(z)=\left(P(z)\right)^c$ for all $z\in X$ when $R$ is complete, therefore $R$ has closed sections and $P$ has open sections, hence $R$ is continuous. 

\medskip

\noindent $\mathbf{\ref{it: c2} \Rightarrow \ref{it: ei2}}$ Theorem \ref{thm: c} implies that the relation is complete and transitive. Then the proof follows from  \citet[Theorem II]{ei41}.  
\medskip

\noindent $\mathbf{\ref{it: c2} \Rightarrow \ref{it: sensitive}}$ Let $R$ be a binary relation $R$ on  topological space $X$ satisfying the assumptions of the hypothesis of the implication and assume it is not sensitive. Then, for all $(x,y)\notin R\cup R^{-1}$ there exists an open neighborhood $V$ of $(x,y)$ such that $V\cap\left( R\cup R^{-1}\right)=\emptyset$. Then $\left( R\cup R^{-1}\right)^c$ is open, hence it has open sections. Note that  $X=P(z)\cup R^{-1}(z)\cup \left( R\cup R^{-1}\right)^c(z)=P^{-1}(z)\cup R(z)\cup \left( R\cup R^{-1}\right)^c(z)$ for any $z\in X$. Then it follows from $P$ has open sections that $R$ has closed sections. Then Theorem \ref{thm: c} implies that $R$ is complete. This furnishes us a contradiction. 
\medskip

\noindent $\mathbf{\ref{it: c2} \Rightarrow \ref{it: ge}}$ is due to \citet[Corollary 3]{ge13},\fn{\citet[Corollary 3]{ge13} showed that this statement is true provided that the space is connected and $R$ is reflexive. However, the statement is true without the reflexivity assumption and the construction in his proof directly follows. Hence, we drop reflexivity.}     $\mathbf{\ref{it: c2} \Rightarrow \ref{it: ch1}}$ is due to \citet[Fundamental Lemma]{ch87} and $\mathbf{\ref{it: c} \Rightarrow \ref{it: ch2}}$ is due to \citet[Theorem]{ch87}.
\medskip

\noindent $ \mathbf{\ref{it: sccont} \Rightarrow \ref{it: c2}}$ Assume $X$ is disconnected. Then there exists an open partition of $X$ consisting of a set $Y$ and its complement $Y^c.$   Define $R= X\times Y^c.$  Then, $P=Y\times Y^c$ is its asymmetric part and $I=Y^c\times Y^c$ its symmetric part. For all $x\in X,$   $R(x)=Y^c$ and  $P(x)=Y^c$ or $P(x)=\emptyset.$ Since $Y^c$ and $\emptyset$ are open, therefore $R$ has closed upper sections and $P$ has open upper sections. The strong non-triviality of $R$ follows from $P(x)=Y^c$ for all $x\in Y$ and $R(x')=Y^c$ for all $x'\in Y^c.$ In order to see that $R$ is transitive, pick $(x,y), (y,z)\in R.$ Then $y,z\in Y^c.$ Hence $(x,z)\in R.$ It follows from $P(x)\cup P^{-1}(y)=Y^c\cup Y$ for all $(x,y)\in P$ that $P$ is negatively transitive. Finally, it is clear that $R$ is incomplete. 
\medskip

\noindent    $\mathbf{\ref{it: ei2} \Rightarrow \ref{it: c2}}$ Assume $X$ is disconnected.  Then there exists an open partition of $X$ consisting of a set $Y$ and its complement $Y^c.$  Let $Q$ be an anti-symmetric,  complete and continuous binary relation on $X$ (since the space is quasi-ordered, such relation exists). Then for all $x\in Y$ and $y\in Y^c,$ either $(x,y)\in Q$ or $(y,x)\in Q.$ Assume without loss of generality that there exist $x\in Y$ and $y\in Y^c$ such that $(y,x)\in Q.$ Define another binary relation $R$ on $X$ as follows. 
\[
\begin{array}{l}
\forall x,y\in Y,  ~ x\in R(y) \Leftrightarrow x\in Q(y),\\ 
\forall x,y\in Y^c,  ~ x\in R(y) \Leftrightarrow x\in Q(y),\\
\forall x\in Y, \forall y\in Y^c,  ~ y\in R(x).\\ 
\end{array}
\] 
  It is clear that $R$ is anti-symmetric and  complete.  
 For all $x\in Y$,  $R(x)=Q(x)\cup Y^c$ and $R^{-1}(x)=Q^{-1}(x)\cap Y.$ For all $y\in Y^c.$  Then $R(y)=Q(y)\cap Y^c$ and $R^{-1}(y)=Q^{-1}(y)\cup Y.$ Since $Q$ has closed sections and $Y, Y^c$ are closed, therefore $R$ has closed sections.  Then, completeness implies $R$ is continuous. 

Note that $Q$ and $R$ has identical ordering both on $Y$ and on $Y^c.$ Moreover, it follows from $(y',x')\in Q$ for some  $(x',y')\in Y\times Y^c$ and $(x,y)\in R$ for all $(x,y)\in Y\times Y^c$ that $Q$ and $R$ have different ordering among the elements of  $Y$ and $Y^c.$ Since $X$ has more than two elements, therefore $Q$ and $R$ are neither identical, nor inverse to each other.
 \medskip

\noindent $\mathbf{\ref{it: ge}, \ref{it: sensitive} \Rightarrow \ref{it: c2}}$  Assume $X$ is disconnected.  Then there exists an open partition of $X$ consisting of a set $Y$ and its complement $Y^c.$  Define $R= Y\times Y^c$ as in the proof of the backward direction implication for $k=1$ in Theorem \ref{thm: c}. Then $R$ is incomplete, non-trivial, transitive, has closed sections and its asymmetric part $P$ has open sections. 
 %
 Now pick an arbitrary pair $(x,y)\in P= Y\times Y^c.$  Since $Y$ and $Y^c$ are open, $Y\times Y^c$ is open. Hence $Y\times Y^c$ is an open neighborhood of $(x,y).$ By construction $P=R=Y\times Y^c.$ Hence $(x', y')\in R$ for all $(x',y')\in Y\times Y^c.$ This furnishes us a contradiction to the fragility of $R.$ This completes the proof of \ref{it: ge}$\Rightarrow$ \ref{it: c2}. Now, pick an arbitrary pair in $(x,y)\in P^c=R^c= \left(Y\times Y\right) \cup \left(Y^c\times Y^c\right)\cup \left(Y^c\times Y\right).$ Since $P^c$ is  open, this furnishes us a contradiction to sensitivity of $R$. This completes the proof of \ref{it: sensitive}$\Rightarrow$ \ref{it: c2}. 
 \medskip

\noindent $\mathbf{\ref{it: ch1} \Rightarrow \ref{it: c2}}$ Assume $X$ is disconnected. Then there exist $Y, Y^c$ non-empty and open subsets of $X.$ Define $P= Y\times Y^c.$ It is clear that $P$ is asymmetric. Define a function $u: X\ra \Re$ as $u(x)=0$ if $x\in Y$ and $u(x)=1$ if $x\in Y^c.$ Since $Y$ and $Y^c$ are open, $u$ is continuous. Moreover $u(x)<u(y)$ if and only if $(x,y)\in Y\times Y^c=P,$ hence $P$ has a continuous dual-representation. We will now show that $P$ is not strongly separable. The relation $R=\{(x,y)~|~(y,x)\notin P\}$ is then defined as $R=(Y\times Y) \cup (Y\times Y^c)\cup (Y^c\times Y^c).$ Pick $(x,y)\in P.$ Then $x\in Y$ and $y\in Y^c. $ By construction of $R,$ for all $x'\in X,$ if $x'\in P(x),$  then $x'\in Y^c.$  Similarly,  for all $y'\in X,$  if $y'\in P^{-1}(y),$  then $y'\in Y.$  Hence $(x',y')\in Y^c\times Y$  for all $x'\in P(x)$ and $y'\in P^{-1}(y).$ It follows from $Y^c\times Y=R^c$ that $(x',y')\notin R$  for all $x'\in P(x)$ and $y'\in P^{-1}(y).$ Therefore $P$ is not strongly separable.
\medskip

\noindent $\mathbf{\ref{it: ch2} \Rightarrow \ref{it: c2}}$ The proof follows from \ref{it: ch1} $\Rightarrow$ \ref{it: c2} above.
\smallskip

The proof of Theorem \ref{thm: cb} is complete.
\prff


We finally turn to the proof of Proposition \ref{thm: discrete}. Before that we need the following notation and a lemma. 
 A binary relation is said to be  an {\it equivalence relation} if it is reflexive, symmetric and transitive. Let $I$ be an equivalence relation on a set $X$, $[x]=\{x'\in X~|~(x,x')\in I\}$ denote the equivalence class of $x$ and the space $X|I$ the {\it quotient space} of $X$ with respect to the relation $I$. Let $\pi: X\ra X|I$ denote the corresponding quotient map defined as $\pi(x)=[x].$  Let $\tau$ be a topology on $X.$ Then the quotient topology is defined as $\hat\tau=\{U\subset X|I ~|~\pi^{-1}(U) \in \tau \}.$  Hence, a set $A$ in $X|I$ is open if and only if $\pi^{-1}(A)$ is open (in $X$). Equivalently,  a set $A$ in $X|I$ is closed if and only if $\pi^{-1}(A)$ is closed.   
For any binary relation $R$ on a set $X$ whose symmetric part $I$ is an equivalence relation, define an induced relation $\hat R$ on $X|I$ as $([x],[y])\in \hat R$ if $(x',y')\in R$ for all $x'\in [x]$ and $y'\in [y].$  Define $\hat I$ as the symmetric part of $\hat R$ and $\hat P$ as its asymmetric part.
%

\lm
Let $R$ be a semi-transitive and continuous binary relation on a topological space $X$ such that  its symmetric part $I$ is an equivalence relation.  Then the induced relation $\hat R$ on $X|I$  is anti-symmetric and continuous. 
\lb{thm: quotient}
\lmm


\prf[\bf Proof of Lemma \ref{thm: quotient}.]   
The semi-transitivity of $R$ implies that $([x],[y])\in \hat R$ if and only if $(x,y)\in R.$ One of the directions is implied by the definition of $\hat R$. In order to prove the other direction, assume $(x,y)\in R.$ Then either $(y,x)\in R$ or $(y,x)\notin R.$ If $(y,x)\in R,$ then $(x,y)\in I.$ Since $I$ is transitive, $(x',y')\in I$ for all $x'\in [x]$ and $y'\in [y].$ If $(y,x)\notin R,$ then $(x,y)\in P.$ Since $R$ is semi-transitive, therefore $(x',y')\in P$ for all $x'\in [x]$ and $y'\in [y].$ Hence, the definitions of $I$ and $P$ imply that $(x',y')\in R$ for all $x'\in [x]$ and $y'\in [y].$ 
 Therefore, $([x],[y])\in \hat P$ if and only if $(x,y)\in P.$  
   The anti-symmetry of $\hat R$ directly follows from  $([x],[y])\in \hat R$ if and only if $(x,y)\in R.$  
Recall that the quotient map $\pi$ is defined as $\pi(x)=[x].$ Then 
$$
\pi^{-1}( \hat R([x]) )= \{y: ([x],[y])\in \hat R  \}=\left\{y: (x,y)\in R \right\}=R(x).
$$
Analogously, $\pi^{-1}( \hat R^{-1}([x]) )=R^{-1}(x),$ $\pi^{-1} ( \hat P([x]) )=P(x)$ and $\pi^{-1} ( \hat P^{-1}([x]) )=P^{-1}(x)$ for all $x\in X.$ Therefore, $\hat R$ has closed sections and $\hat P$ has open sections. 
\prff

\prf[\bf Proof of Proposition \ref{thm: discrete}]
Assume $X$ is topological space and $R$ is a complete, semi-transitive and continuous binary relation on it. It follows from \citet[Theorem I]{se69} that its symmetric part $I$ is transitive. Define a binary relation $\hat R$ on the quotient space $X| I$  as $([x], [y])\in \hat R$ if and only if $(x',y')\in R$ for all $x'\in [x]$ and all $y'\in [y]$. The definition of the induced relation $\hat R$ on the quotient space implies its completeness.  It follows from Lemma \ref{thm: quotient} that $\hat R$ is anti-symmetric and continuous. Pick $[x],[y]\in X|I$ such that $[x]\neq [y].$ Assume without loss of generality that $[x]\in \hat P([y]).$ If there exists $z\in X$ such that $[z]\in \hat P([y])\cap \hat P^{-1}([x]),$ then $[x]\in \hat P([z])$ and $y\in P^{-1}(z).$ Hence, $\hat P([z])$ and $\hat P^{-1}([z])$ are disjoint and open neighborhoods of $[x]$ and $[y],$ respectively. If $\hat P([y])\cap \hat P^{-1}([x])=\emptyset,$  then $[x]\in \hat P([y])$ and $[y]\in \hat P^{-1}([x])$ imply  $\hat P([y])$ and $\hat P^{-1}([x])$ are disjoint and open neighborhoods of $[x]$ and $[y],$ respectively.\fn{A similar result for anti-symmetric binary relations is provided by \citet[1.4]{ei41}.} 
\prff

\bigskip

\setlength{\bibsep}{5pt}
\setstretch{1}


\bibliographystyle{econometrica} 
\bibliography{References.bib}

\end{document}